\documentclass[10pt, conference, compsocconf]{IEEEtran}

\usepackage{mathptmx}
\usepackage{fancyhdr}
\usepackage[normalem]{ulem}
\usepackage[hyphens]{url}
\usepackage[sort,nocompress]{cite}
\usepackage[final]{microtype}
\usepackage[keeplastbox]{flushend}
\usepackage{soul}
\usepackage{algorithm}
\usepackage{xspace}
\usepackage[font=small,labelfont=bf]{caption}
\usepackage{mathtools}
\usepackage{xcolor}
\usepackage{amsfonts}
\usepackage{amssymb}
\usepackage{pifont}
\usepackage{color}
\usepackage{algpseudocode}
\usepackage{multirow}
\usepackage{textgreek}
\usepackage{subfig}
\usepackage{textcomp}
\usepackage[]{siunitx}


\usepackage[bookmarks=true,breaklinks=true,letterpaper=true,colorlinks,linkcolor=black,citecolor=blue,urlcolor=black]{hyperref}

\newcommand{\old}[1]{}
\newcommand{\fig}[1]{Figure~\ref{#1}}
\newcommand{\sect}[1]{Section~\ref{#1}}
\newcommand{\tab}[1]{Table~\ref{#1}}
\newcommand{\algo}[1]{Algorithm~\ref{#1}}
\newcommand{\eqn}[1]{Equation~\ref{#1}}

\newcommand{\proposed}[0]{DiVa\xspace}

\algnewcommand{\LineComment}[1]{\State \(\triangleright\) #1}
\newcommand{\x}[0]{\texttt{X}\xspace}
\newcommand{\y}[0]{\texttt{Y}\xspace}
\newcommand{\w}[0]{\texttt{W}\xspace}
\newcommand{\gx}[0]{\texttt{G(X)}\xspace}
\newcommand{\gw}[0]{\texttt{G(W)}\xspace}
\newcommand{\reweight}[0]{DP-SGD(R)\xspace}

\newcommand\blfootnote[1]{%
\begingroup
\renewcommand\thefootnote{}\footnote{#1}%
\addtocounter{footnote}{-1}%
\endgroup
}

\begin{document}
\title{\LARGE DiVa: An Accelerator for Differentially Private Machine Learning}

\author{
\IEEEauthorblockN{
    Beomsik Park\IEEEauthorrefmark{1}\qquad
    Ranggi Hwang\IEEEauthorrefmark{1}\qquad
    Dongho Yoon\qquad 
    Yoonhyuk Choi\qquad
    Minsoo Rhu}

\IEEEauthorblockN{
    School of Electrical Engineering\\
    KAIST\\
    \{\texttt{parkbeomsik, ranggi.hwang, dongho.yoon, yoonhyuk.choi, mrhu\}@kaist.ac.kr}
    }

}

\maketitle

\begin{abstract}

The widespread deployment of machine learning (ML) is raising
serious concerns on protecting the privacy of users who contributed to
the collection of training data.  Differential privacy (DP) is rapidly
gaining momentum in the industry as a practical standard for privacy
protection.
Despite DP's importance, however,
				little has been explored within the computer systems community
				regarding the implication of this emerging ML
				algorithm on system designs. In this work, we conduct a detailed
				workload characterization on a state-of-the-art differentially
				private ML training algorithm named DP-SGD.  We uncover
				several unique properties of DP-SGD (e.g., its high memory capacity and computation requirements vs.
						non-private ML), root-causing its key bottlenecks. Based on our
				analysis, we propose an accelerator for differentially private ML named
				\proposed, which provides a significant improvement in
				compute utilization, leading to $2.6\times$ higher
				energy-efficiency vs. conventional systolic arrays.
	\end{abstract}


\IEEEpeerreviewmaketitle
\blfootnote{
\\
\IEEEauthorrefmark{1} Co-first authors who contributed equally to this research.
}

\section{Introduction}

Deep neural network (DNN) based machine learning (ML) algorithms have
demonstrated remarkable performance in numerous application domains~\cite{resnet,bert,alphago}.
Such advances are 
	fueled by the availability of large, representative datasets that are being
	utilized for training ML algorithms, allowing DNNs to capture multi-level
	representations and abstractions from the training dataset.

Despite the enormous success of DNNs, the widespread deployment of ML 
applications is raising serious concerns on protecting the \emph{privacy} of
users who contributed to the collection of training data. Because the training
datasets are oftentimes crowdsourced and can include sensitive information
(e.g., private emails, medical records, financial transactions), an adversary
can seek a training data extraction attack to restore individual training
examples.  Even if the ML model parameters are not shared,
	black-box access to the models was shown to  leak private
	information~\cite{fredrikson2015model,shokri2017membership,nicholas2021extracting,hayes2010logan}.  In particular, recent literature demonstrated
	that the memorization behavior exhibited with large DNN models can be
	exploited to leak individual's private information, posing risks to those that
	contributed to the training data. 

Given this landscape, both industry and academia have started developing
solutions that satisfy the demands of ML applications while also offering
principled and rigorous privacy guarantees~\cite{dpsgd,bassily2014private,pate,shokri2015ppdl,dong2017dropping}. Among various
privacy mechanisms, \emph{differential privacy} (DP)~\cite{dwork2006dp} has rapidly
gained momentum as a well accepted notion of privacy (\sect{sect:background_dp}). 
Informally speaking, for
the ML model to be differentially private, the estimated model and all of its
parameters should be \emph{indistinguishable} regardless of whether a particular client's
data was taken into consideration or not during the training process, allowing
the privacy of \emph{individual} training examples to be protected.  
Thanks to
DP's strong mathematical guarantees on privacy protection and recent advances
in successfully training differentially private ML models~\cite{li2022large, anil2021large,yu2022differentially,hoory2021learning}, we
are witnessing a wide variety of real-world products incorporating DP. For example, Apple
		employs DP to collect iOS device users' anonymous usage patterns and Amazon
		similarly utilizes DP to access user's personalized shopping preference
		while hiding sensitive information regarding past
		purchases~\cite{diethe2020preserving,differential2017learning}. In general, DP methods are being recognized in the
industry as a practical standard for privacy protection.  Despite DP's
importance, however, little has been explored and understood within the
computer architecture community regarding the implication of this
highly important and emerging ML algorithm on computer system designs.

Consequently, an important motivation and {\bf key contribution} of our study
is a detailed workload characterization on differentially private ML.  To the
best of our knowledge, this work is the first to quantitatively analyze a
representative, state-of-the-art differentially private ML system, discussing its
architectural implications as well as its key challenges.  Standard practice in
training a \emph{non-private} DNN model is to employ stochastic gradient
descent (SGD) where multiple input examples are \emph{batched} together as a
training mini-batch. During backpropagation, SGD derives	a \emph{per-batch}
weight gradient for updating the DNN weights.  Under the
privacy-preserving scenario, the state-of-the-art algorithm employed in
practice is the ``differentially-private'' SGD (henceforth referred to as
		DP-SGD), a variation of SGD with strong privacy protection.  The unique
property of DP-SGD is twofold: 1) it requires the derivation of
\emph{per-example} weight gradients, rather than the \emph{per-batch} weight
gradients derived in non-private SGD, and 2) these per-example weight
gradients go through series of \emph{post-processing} steps (e.g.,
		per-example gradient norm derivation, gradient clipping/reduction, and
		random noise addition to the gradients)
	for the privacy enhancement of the target model (\sect{sect:background_dpsgd}).

	Given such, our study uncovers important research challenges this emerging privacy enhanced ML
	paradigm brings about to computer system designers, which we outline below.
	The key compute primitive of SGD training is the GEMM (generalized matrix
			multiplication) operation as it can cover the majority of execution time
	of both forward and backpropagation
	(\sect{sect:char_tpu_training_time_breakdown}).  Systolic arrays are arguably
	the most successfully deployed GEMM acceleration engine purposed for ML
	training, notably represented by its wide adoption in the industry
	(e.g., Google TPUs~\cite{jouppi2020domain}).  Using Google Cloud TPUv3, we demonstrate
	that training DNNs with DP-SGD incurs both low compute utilization (up
			to $29\times$ lower than SGD) and high memory consumption (up to
				$11\times$ higher than SGD), significantly aggravating training
			throughput by up to $33\times$ vs. non-private SGD.  Careful
			examination of such performance drop reveals that conventional systolic
			arrays are suboptimal for efficiently handling both the derivation of
			per-example weight gradients as well as gradient post-processing steps,
			posing serious challenges in practically training differentially private,
			large-scale DNN models (\sect{sect:characterization_compute}).

To this end, we present {\bf \proposed}, an accelerator architecture tailored
for the unique algorithmic properties of \underline{{\bf Di}}fferentially
Pri\underline{{\bf Va}}te machine learning training. The design of \proposed is
driven by our detailed characterization study, unlocking DP-SGD's full potential
with the following two key innovations.

\begin{enumerate}

\item Compared to non-private SGD, deriving DP-SGD's per-example weight
gradients can entail hundreds of irregular, tall-skinny shaped GEMMs,
					which is ill-suited for systolic arrays optimized for
					regular, square shaped GEMMs. We propose an \emph{outer-product}
					based dataflow for DP-SGD's processing engines (PEs) which provides
					high robustness to both regular and irregularly shaped GEMMs.
					Compared to systolic array's dataflow, our
					proposal does a much superior job in mapping irregular
					(per-example weight gradient) GEMMs over the compute fabric, significantly
					improving PE utilization by an average
					$5.5\times$.

\item Another critical step in DP-SGD training is the gradient post-processing
stage where the L2 norm values of per-example gradients are derived, which are
utilized for clipping and reducing the gradients
(\sect{sect:background_dpsgd}).  All of these operations are highly memory
bandwidth limited and can cause  latency overheads.  \proposed augments its
outer-product based PE array with a tightly coupled DP-SGD
\emph{post-processing unit} (PPU), a \emph{multi-level adder-tree} for vector
reductions whose datapath is optimized for the unique dataflow of gradient
norm/clipping/reduction to minimize off-chip memory accesses. We demonstrate
that \proposed's PPU provides $99\%$ reduction in off-chip data
movements during gradient post-processing, effectively resolving its memory
bandwidth limitation.

			\end{enumerate}

Putting everything together, \proposed provides a significant improvement
in compute utilization which leads to an average $3.8\times$ training time reduction,
providing $2.6\times$ higher energy-efficiency vs. conventional systolic arrays for DP-SGD.

\section{Background}
\label{sect:background}

\subsection{Why Differential Privacy?}
\label{sect:background_dp}

As ML become widely deployed across various application domains, the importance
of protecting data privacy is growing rapidly, especially among areas such as
finance, health care, etc~\cite{blatt2020secure,heaton2017deep,jakob2016learning}.  \emph{Differential privacy} (DP) has
recently emerged as a privacy preserving mechanism that provides a strong,
				 mathematical definition of privacy under the context of statistical
				 and ML analysis.  In general, an algorithm is considered
				 differentially private if an observer seeing the output of the
				 algorithm cannot tell whether a particular individual's information
				 was utilized in computing that given output, 
allowing the privacy of individual training examples to be protected.
				 Overall, DP mathematically guarantees that the observer seeing the
				 output of a given algorithm will make the same inference about any
				 individual's private information, regardless of whether or not that
				 individual's information is included in the input.
				 We refer to
				 \cite{dwork2006dp,dpsgd,dwork2014algorithmic,dwork2008differential} for a more rigorous discussion on DP's mathematical
				 foundation.

\begin{figure}[t!] \centering
\includegraphics[width=0.475\textwidth]{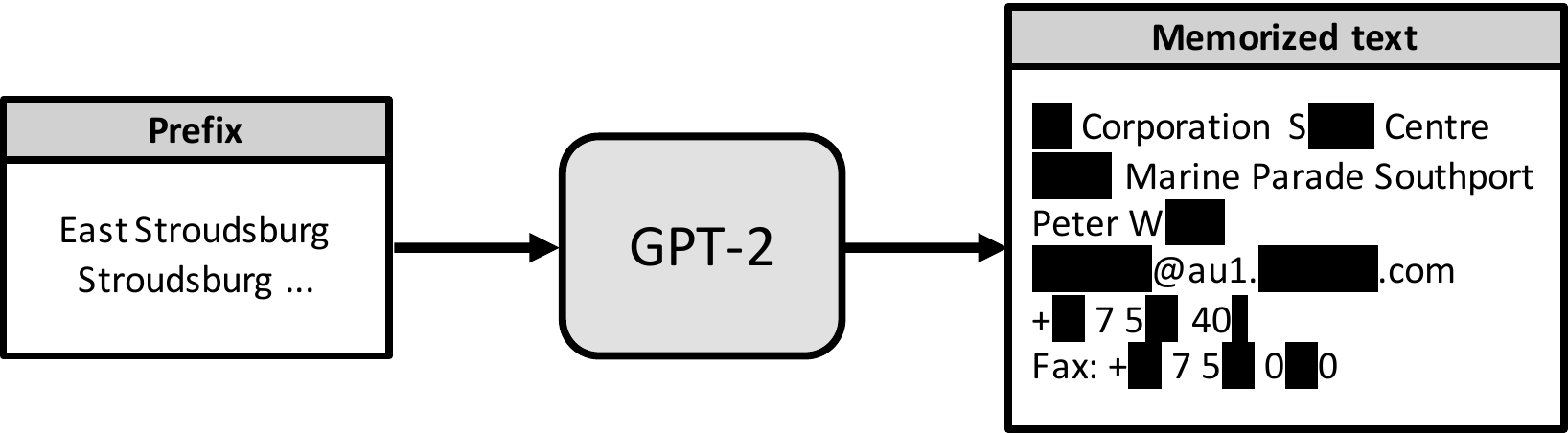}
\vspace{0.5em}
\caption{
	When one prompts the GPT language model with the prefix ``East Stroudsburg Stroudsburg $\ldots$'', GPT was shown to autocomplete a block of text that contains the
		full name, phone number, e-mail address, and physical address of a particular
		individual whose information was included in GPT's training dataset. The figure
	is reproduced from \cite{nicholas2020blog}.
}
		\vspace{-0.5em}
\label{fig:privacy_leak_example}
\end{figure}

				 Under the context of ML, DNN models trained with an individual's data
				 (e.g., clinical records, photos) were shown to be vulnerable to
				 attacks that directly analyze the internal model parameters or
				 indirectly query the model repeatedly in a black-box setting
				 (\fig{fig:privacy_leak_example})~\cite{fredrikson2015model,shokri2017membership,hayes2010logan,nasr2019comprehensive}.  What is troubling is
				 the fact that larger DNN models are more vulnerable than smaller
				 ones~\cite{nicholas2021extracting}, which is at odds with recent trends where larger/bigger
				 models are favored given their higher algorithmic
				 performance~\cite{gpt2,brown2020gpt3,shazeer2017moe,smith2022using}.  To
				 address such vulnerabilities, the seminal work by Abadi et
				 al.~\cite{dpsgd} proposed a solution that enables the training of
				 DNNs with DP. In the remainder of this section, we review both a
				 non-private SGD vs. privacy-aware DP-SGD and discuss their key
				 differences.

\subsection{Non-Private Training with SGD}
\label{sect:background_sgd}

Training a DNN involves learning and
		 updating the \emph{weights} of the DNN layers by the
		 operations of forward and backward propagation (aka backpropagation)
	as detailed below.

	\begin{figure}[t!] \centering
    \subfloat[]
    {
\includegraphics[width=0.485\textwidth]{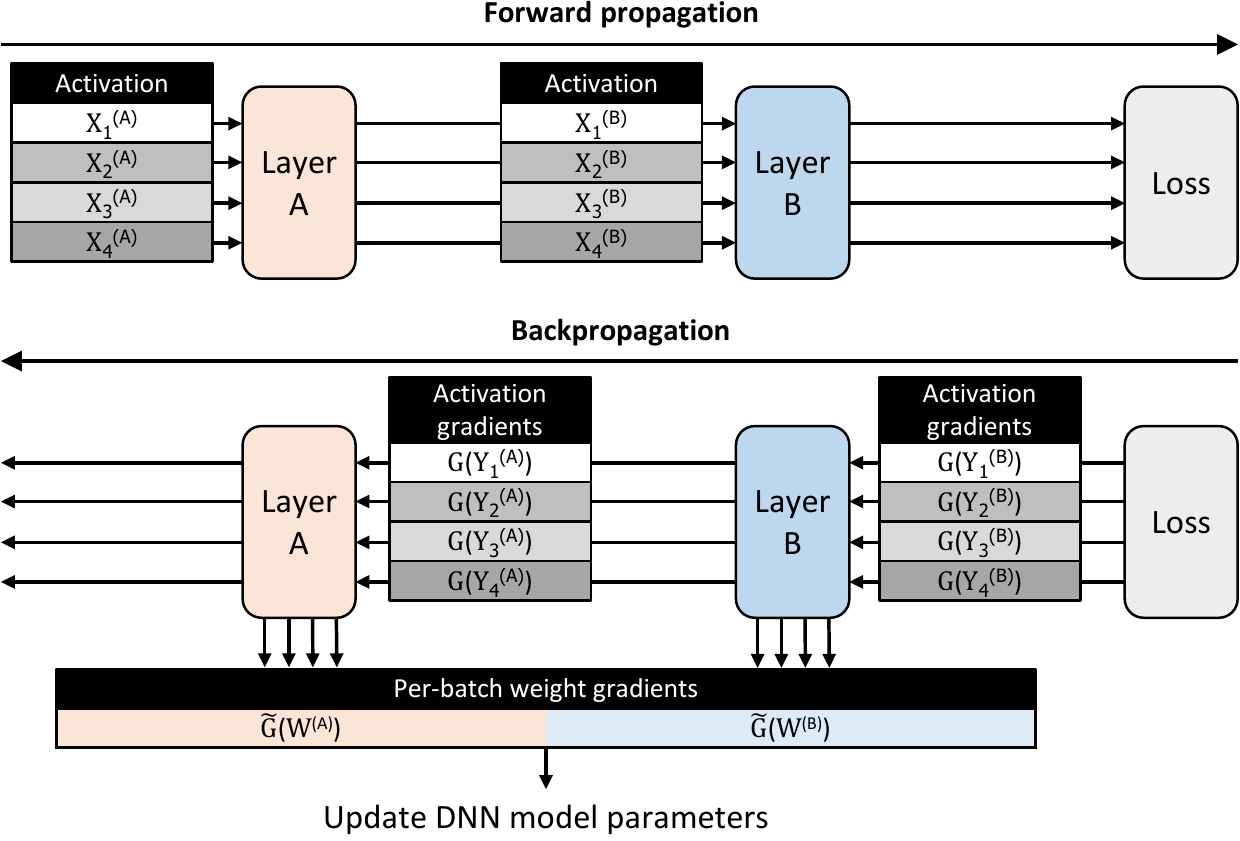}
    }
    \vspace{-0em}
    \subfloat[]
    {
 \includegraphics[width=0.485\textwidth]{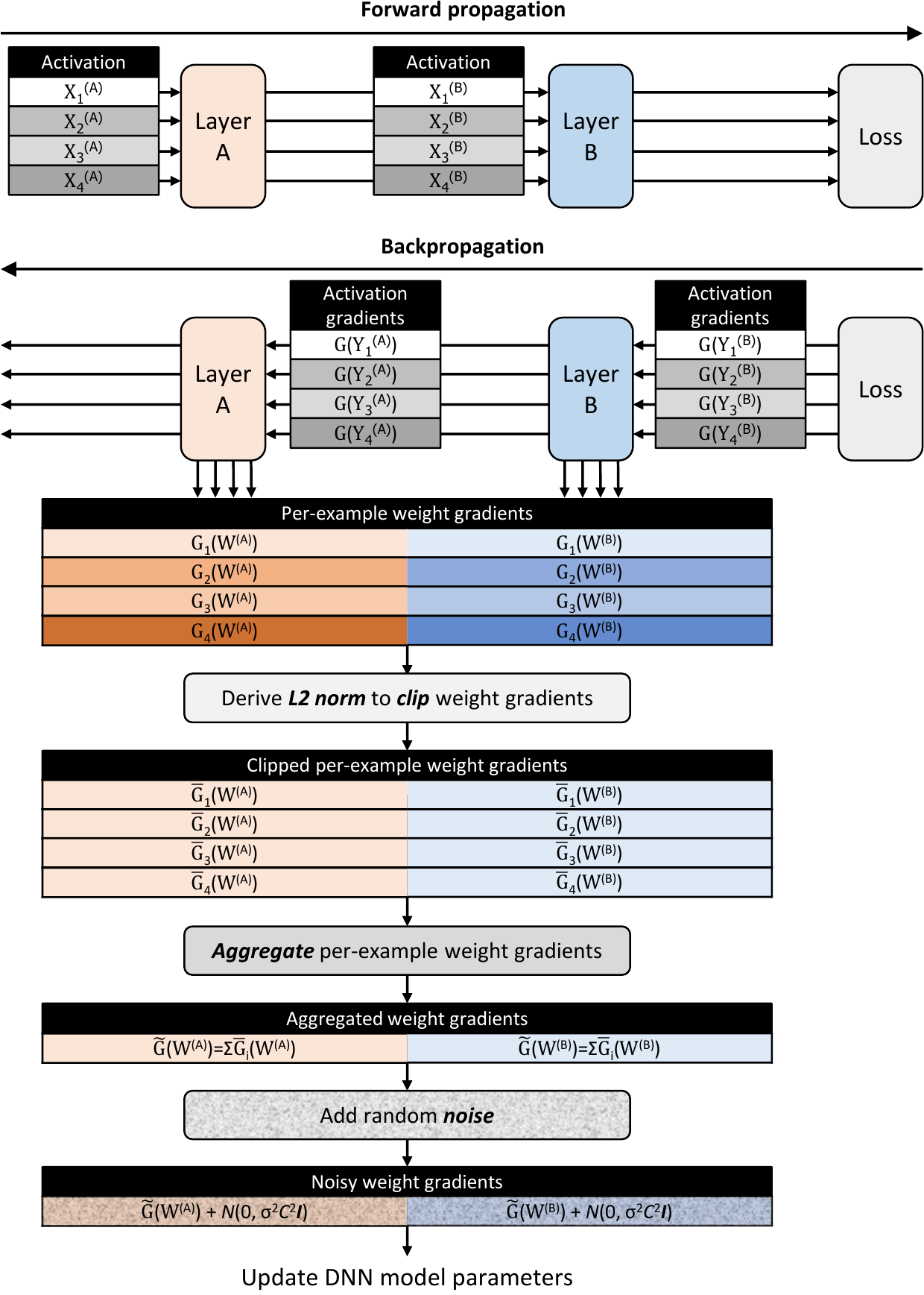}
    }
    \vspace{0em}
\caption{Training with a (a) non-private SGD and (b) DP-SGD. Example assumes $Y$=$X\times$$W$ and an input mini-batch
size of $4$. Both layers are assumed to have its corresponding weight values, necessitating the derivation of per-layer weight gradients.
As depicted, DP-SGD requires $4$ times larger memory allocations than SGD for storing its per-example weight gradients (\gw).}
    \label{fig:overview_sgd}
		\vspace{-1.3em}
\end{figure}

{\bf Forward propagation.} \fig{fig:overview_sgd}(a) illustrates the forward
propagation of a two-layered feedforward DNN with a mini-batch
size of $4$.  Each layer conducts a mathematical operation (e.g., convolution)
	to its input activation (\x) using the per-layer weight (\w), if any, and
	generates the output activation (\y).  Note that all
	the (four) input examples that are part of the mini-batch go through the
	layer-wise forward propagation \emph{in parallel}, which helps better reuse
	the weight \w and improve compute utilization of the ML accelerator. 

{\bf Backpropagation.} A \emph{loss function} is used to derive the magnitude
of an inference's error at the end of forward propagation. Specifically, the
\emph{gradient} of the loss function with respect to the last layer's input
activation is derived. Using the \emph{chain rule}, all the layer's input
activation gradient (\gx) as well as the weight gradient (\gw) is calculated
on  a per-layer basis, from the last layer to the first layer~\cite{lecun_gd,rhu:2016:vdnn}.
Once the per-layer \gw is derived, it is utilized to update the
corresponding layer's weight \w for training. A distinguishing aspect of
mini-batch SGD training is that the size and shape of the per-layer \gw 
is \emph{identical} to the original per-layer weight \w, regardless of the
mini-batch size. This is because the weight gradient vectors derived for the
individual input examples (that constitute the input mini-batch) are \emph{aggregated}
(i.e., reduced) into a single set of \gw. This is illustrated in
\fig{fig:overview_sgd}(a) where only a single set of \gw is derived, per-layer. In
the remainder of this paper, we refer to the non-private SGD's weight gradients
as ``per-batch'' weight gradients to distinguish it against the ``per-example''
		weight gradients of DP-SGD, as detailed below.

\begin{algorithm}[t!]
\caption{DP-SGD and DP-SGD(R)}
\label{algo:dpsgd}
\renewcommand{\algorithmicrequire}{\textbf{Input:}}
\renewcommand{\algorithmicensure}{\textbf{Output:}}
\begin{algorithmic}[1]
\scriptsize
\Require Dataset $D=\{ \left( x_1,y_1 \right), \ldots ,  \left( x_N,y_N \right)\}$,
			batch size $B$, max gradient norm $C$, noise multiplier $\sigma$, max steps $T$, learning rate $\eta_t$, loss function $l$, model weight $w=\{w^{(1)}, w^{(2)}, \ldots, w^{(M)}\}$, gaussian random varaible $\boldsymbol{N}$
\State \textbf{Initialize} model weight $w_0$
\For{$t = 0, 1, \ldots , T$}
	\State $\textrm{Randomly sample a minibatch $\{(x_i, y_i) \mid i \in [B]\}$ from dataset $D$}$
	\State \LineComment{Compute loss value through \textit{forward propagation}}
	\State \textit{For each $i \in [B]$, } $L_i \gets l(w_t, x_i, y_i)$
	\State \LineComment{Derive \emph{differentially private}  weight gradients}
	\State $g\sp{\prime}_t \gets$ \textsc {Derive\_DP\_Gradients($L$) \\ \qquad\qquad\qquad or Derive\_Reweighted\_DP\_Gradients($L$)}
	\\
	\State $w_{t+1} \gets w_t - \eta_tg\sp{\prime}_t$ \Comment{\textit{Update} model weight}
\EndFor
\\
\LineComment{DP-SGD}
\Procedure {Derive\_DP\_Gradients($L$)}{}
\LineComment{Compute \textit{per-example} weight gradients through \textit{backpropagation}}
\For{$\textit{layer m} = M, M-1, \ldots, 1$}
		\State \textit{For each $i \in [B]$, } $g_i(w^{(m)}) \gets \frac{\partial L_i}{\partial w^{(m)}}$
\EndFor
\\
\State \textit{For each $i \in [B]$, } $n_i \gets \Vert g_i(w) \Vert_2$ \Comment{Derive per-example \textit{L2 norm}}
\State \textit{For each $i \in [B]$, } $\bar{g}_i(w) \gets g_i(w)/\textrm{max}(1, \frac{n_i}{C})$ \Comment{\textit{Clip} per-example gradients}
\State \Return $\frac{1}{B}\left( \sum_{i\in [B]}^{} \bar{g}_i(w) + \boldsymbol{N}(0, \sigma^2C^2\textbf{I}) \right)$ \Comment{\textit{Aggregate} gradients \& add \textit{noise}}
\EndProcedure
\\
\LineComment{DP-SGD(R)}
\Procedure {Derive\_Reweighted\_DP\_Gradients($L$)}{}
\LineComment{Compute \textit{per-example} weight gradient \textit{L2 norm} via \textit{1st backpropagation}}
\For{$\textit{layer m} = M, M-1, \ldots, 1$}
		\State \textit{For each $i \in [B]$, } $n^{(m)}_i \gets \Vert \frac{\partial L_i}{\partial w^{(m)}} \Vert_2$
\EndFor
\State \textit{For each $i \in [B]$, } $n_i = \Vert \{n^{(1)}_i, n^{(2)}_i, \ldots , n^{(M)}_i\} \Vert_2$
\\
\State $L\sp{\prime} \gets \sum_{i\in [B]}^{} L_i/\textrm{max}(1, \frac{n_i}{C})$ \Comment{Compute \textit{reweighted} loss value}
\State \LineComment{}{Compute \textit{clipped, per-batch} weight gradient via \textit{2nd backpropagation}}
\For{$\textit{layer m} = M, M-1, \ldots, 1$}
	\State $\tilde{g}(w^{(m)}) \gets \frac{L\sp{\prime}}{\partial w^{(m)}}$
\EndFor
\State \Return $\frac{1}{B}\left( \tilde{g}(w) + \boldsymbol{N}(0, \sigma^2C^2\textbf{I}) \right)$ \Comment{Add random \textit{noise}}
\EndProcedure
\\
\Ensure Differentially private model weight $w_T$ and total privacy cost $(\epsilon, \delta)$
\end{algorithmic}
\end{algorithm}

\subsection{Privacy-Aware Training with DP-SGD}
\label{sect:background_dpsgd}

{\bf High-level overview of DP-SGD.} Abadi et al.~\cite{dpsgd} suggests to
add DP to deep learning models by adding bias and noise into the mini-batch
gradient computation process.  \algo{algo:dpsgd} provides a high-level overview
of such DP-SGD training procedure.  At each step of the training iteration,
	 DP-SGD derives the weight gradients for \emph{each individual input
		 examples}  that constitute the input mini-batch (line $19$), rather than
		 computing a single set of weight gradient per each mini-batch as done in
		 SGD (see \fig{fig:overview_sgd}).  These \emph{per-example} weight
		 gradients are then \emph{clipped} (line $23$) based on the L2 norm of each individual
		 per-example gradient (line $22$) and subsequently \emph{reduced} into a
		 single set of weight gradient \gw (line $24$). The reduced \gw is then added
		 with noise to protect privacy (line $24$).  By taking a step in the opposite
		 direction of this noisy gradient, the DNN model is incrementally trained
		 in a differentially private manner.

{\bf SGD vs. DP-SGD.} Compared to SGD, a distinguishing aspect of
DP-SGD is threefold. First, DP-SGD requires the derivation of per-example
weight gradients rather than per-``batch'' weight gradient of SGD.  Second,
			 derivation of per-example weight gradients requires \emph{separate}
			 memory allocations for each, per-example weight gradient \gw across
			 all the layers as it is needed for computing per-example L2 norms (line $22$),
incurring a significant increase in memory usage, i.e., compared to SGD
			 which requires \texttt{sizeof}(\gw) memory allocation, DP-SGD with
			 mini-batch of size $B$ requires $B$$\times$\texttt{sizeof}(\gw) memory
			 allocation per each layer (\fig{fig:overview_sgd}).  Lastly, the per-example weight
			 gradients require post-processing (i.e., gradient norm
					 derivation, gradient clipping, gradient reduction, and noise
					 addition) in order to derive the single set of \gw to be utilized
			 for model updates. 

{\bf ``Reweighted'' gradients for memory-efficient DP-SGD.} As we detail in
\sect{sect:characterization_memory}, DP-SGD's high memory consumption limits
the maximum mini-batch size that can practically be employed for training,
		posing yet another challenge in deploying DP-SGD.  To address such memory
		allocation problem of DP-SGD, Lee et al.~\cite{lee2021scaling} proposed an optimization
		named \emph{reweighted} DP-SGD (henceforth referred to as DP-SGD(R)) which
		helps reduce the memory allocation size at the cost of additional
		computation steps. Line $28-42$ in \algo{algo:dpsgd} summarizes the steps
		undertaken in DP-SGD(R)'s backpropagation. A key difference
		between DP-SGD vs. DP-SGD(R) is threefold. First, DP-SGD(R) effectively executes
		backpropagation ``twice'' for: 1) deriving \emph{per-example} weight
		gradients during the 1st backpropagation to compute the L2 norm (line
				$31$), and 2) utilizing the L2 norms derived to compute
				\emph{per-batch} weight gradients during the 2nd backpropagation pass (line $39$).
				Second, the gradient clipping and reduction stages of DP-SGD are all \emph{fused}
				as part of DP-SGD's 2nd backpropagation stages (line $39$).
				Third, because per-example weight gradients are only required to compute per-example L2 norms (line $22$)
	and such procedure is fused as part of the 1st backpropagation under DP-SGD(R), the runtime memory manager need not have
	to \emph{overprovision} the memory allocation size with
 $B\times$\texttt{sizeof}(\gw) across all the layers, enabling opportunities to reduce memory usage.
	In \sect{sect:characterization}, we provide a detailed analysis on the compute vs. memory usage tradeoffs between DP-SGD vs. DP-SGD(R).

\begin{figure*}[t!] \centering
\includegraphics[width=0.99\textwidth]{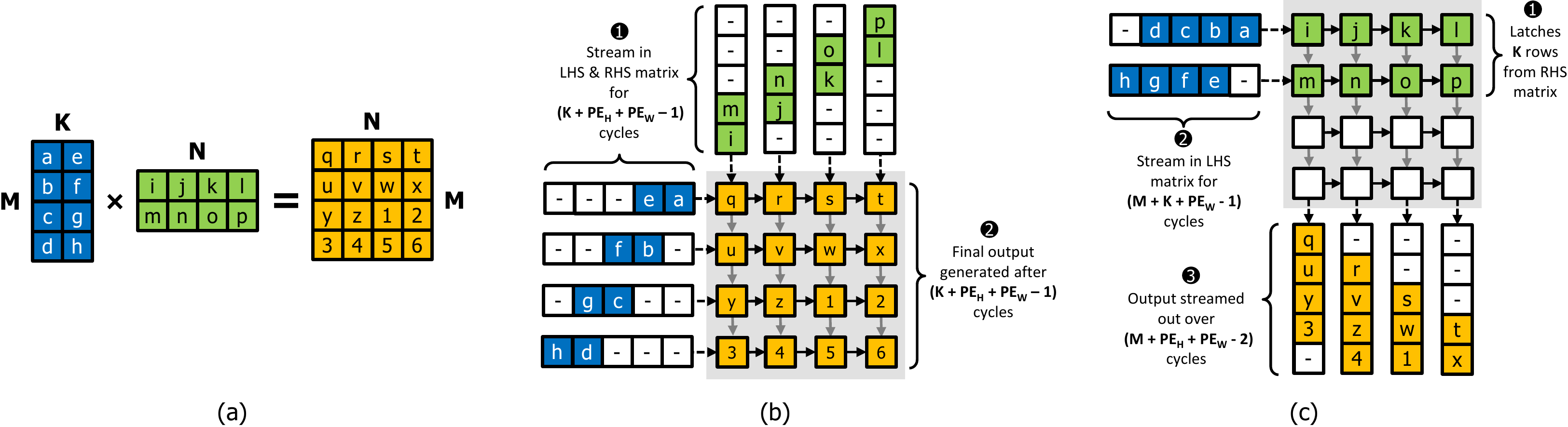}
\vspace{-0.3em}
\caption{(a) The three GEMM dimensions (M,K,N) that define a matrix multiplication operation between (M,K)$=$($4\times2$) and (K,N)$=$($2\times4$) matrices, generating the output matrix (M,N)$=$($4\times4$). Figure (b-c) shows the schematic illustrating two representative dataflows in a systolic array: (b) output stationary (OS) and (c) weight stationary (WS). The height (PE$_{H}$) and width (PE$_{W}$) of the systolic array are assumed as ($4$,$4$) in the given example.
}
\vspace{-0.5em}
\label{fig:systolic_dataflow}
\end{figure*}

\subsection{Systolic Arrays for Accelerating GEMM}
\label{sect:background_systolic}

{\bf Why systolic arrays for training?} There is a rich set of prior literature
on designing accelerator designs for both ML
training and
inference~\cite{cnvlutin,diannao,dadiannao,eyeriss,scnn,song:2015:eie,tpu1,jouppi2020domain,jouppi2021ten,neurocube,maeri,cambricon,rhu:2018:cdma,centaur,tensordimm,tensorcasting,trim,kim2020trim,mcdla,mcdla:cal,kwon:2019:disagg,neummu}.
Interestingly, while defining a \emph{generic} ML accelerator for inference is
challenging (i.e., accelerators for inference is typically optimized for a
		specific application domain, for instance convolution for computer
		vision~\cite{eyeriss,dadiannao}), those for training have more or less
settled on a design that is optimized for a ``single'' key primitive:
generalized matrix multiplication (GEMM). A key reason why accelerators for
training are optimized for GEMM is because both forward and backpropagation of
SGD (i.e., derivation of \gx and \gw) can all be
\emph{permuted} to GEMM for representative DNN layers (e.g., the \emph{im2col}
		operation that transforms convolutions into GEMM~\cite{chetlur2014cudnn,jia2014learning}). Among these,
	systolic arrays have been most commercially successful thanks to its regular
	layout of processing engines (PEs), efficient inter-PE communication, and
	highly regular dataflow with decent data reuse, enabling low power
	consumption and high throughput for training. In the rest of this paper,
	we assume systolic arrays  as the baseline accelerator for
	training given its enormous industrial success and wide applicability.

{\bf Systolic array dataflow.} There are two distinct approaches in mapping the
GEMM's dataflow onto systolic arrays, namely \emph{output stationary}
(OS) and \emph{weight stationary} (WS). The OS dataflow, as depicted in
\fig{fig:systolic_dataflow}(b), refers to the mapping strategy where each PE is
responsible for conducting all the computations required for deriving a given
output activation. All the required data operands are streamed in from the
(left and top) edges of the array, which are distributed to the PEs using local
communication channels to the systolic array. The partial sums are 
generated and reduced down to its final output activation value \emph{locally}
within each PE. Once all the PEs within the systolic array are done deriving
its share of the output activation value, the inter-PE communication links are
utilized to transfer the final outputs out of the array.

	Unlike the OS dataflow, the WS dataflow employs a different strategy as shown
	in \fig{fig:systolic_dataflow}(c). Here, the  weight values (RHS matrix) are
	filled into the local latches in each PE \emph{in advance}, prior to the
	start of GEMM.  The elements of the input activation (LHS matrix)
	are then streamed in through the (left) edge of the systolic array, where
	each PE computes one partial sum every cycle. The partial sums derived are
	then reduced across the rows, along each column in parallel to generate one
	output value per each column. 	Google TPUs are well-known to employ a WS
	dataflow because of its cost-effective design and lower on-chip data fetch
	bandwidth requirements~\cite{norrie2020google,tpu1,jouppi2020domain}. In this
	work, we assume a WS dataflow for our baseline systolic array. Nonetheless,
	we discuss the implication of DP-SGD on OS in
	\sect{sect:proposed_ppu} for the completeness of our study.

In general, because systolic arrays are designed as one large, inflexibly 2D
array ((PE$_{H}$,PE$_{W}$)=($128$,$128$) in Google TPUv3), it can suffer from
low PE utilization when the GEMM's (M,K,N) dimensions do not align with the
dimensions of the \emph{physical} systolic array (e.g., non-square shaped
		matrices, GEMMs with small K-dimension sizes).  For instance, GEMMs with
small K-dimensions map poorly to OS with large (PE$_{H}$, PE$_{W}$)
	systolic array because the two input vectors streaming in from left/top edges
	lead to significant idle cycles along the diagonal direction
	(\fig{fig:systolic_dataflow}(b)). WS similarly suffers from low PE
	utility as it fails in fully utilizing the PEs throughout the computation,
	i.e., only half of the PE rows in \fig{fig:systolic_dataflow}(c) are latched
	with the RHS matrix, reducing effective PE throughput.

\section{Workload Characterization}
\label{sect:characterization}

In this section, we utilize Google Cloud TPUv3~\cite{cloud_tpu} combined with our
cycle-level simulation framework to conduct a workload characterization  on
training representative DNNs via 1) non-private SGD, 2) DP-SGD as-is and 3)
DP-SGD \emph{with} reweighting for memory optimization (denoted \reweight, see
		\sect{sect:background_dpsgd}).  \sect{sect:methodology} further details our
methodology regarding hardware/software configurations, simulation framework,
						benchmark selection, etc.

\subsection{DP-SGD's Memory Consumption and Its Effect on Training Mini-batch Size}
\label{sect:characterization_memory}

\begin{figure}[t!] \centering
\includegraphics[width=0.485\textwidth]{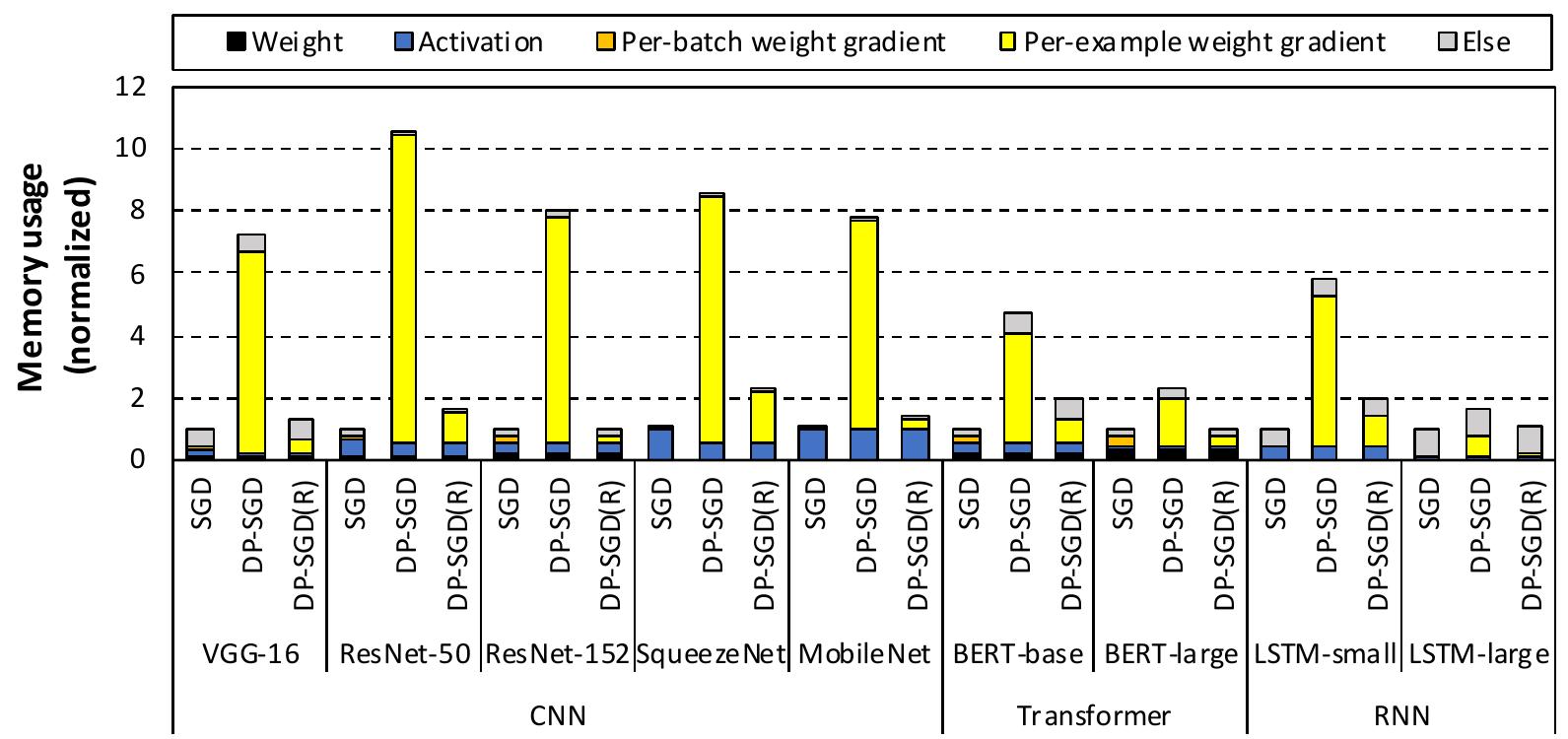}
\caption{Breakdown of SGD, DP-SGD, and \reweight's Google TPUv3 memory usage 
	based on its functionality (normalized to SGD). 
	All three design points assume an identical mini-batch size, i.e., the mini-batch that can be
		employed with DP-SGD is the smallest (vs. SGD and \reweight) so all three designs employ
		such mini-batch size identically for a fair comparison. 
}
\vspace{-0.5em}
\label{fig:char_memory_usage}
\end{figure}

\fig{fig:char_memory_usage} shows the size of memory allocations based on its
functionality.  Training with DP-SGD requires the derivation of per-example
weight gradients whose size grows proportional to the mini-batch size
(\sect{sect:background_dpsgd}). This is represented by the large portion of
per-example weight gradients' memory allocations in DP-SGD, amounting to an
average $78\%$ of its memory consumption. The implication of DP-SGD's
high memory usage is that the maximum possible mini-batch it can practically
employ is severely limited because TPUs come with much smaller
memory size than CPUs (e.g., $16$ GB in Google TPUv3 vs. several TBs in
		CPUs).  For instance, while SGD can train ResNet-152 and BERT-base
with a mini-batch size of $8192$ and $1024$ respectively, DP-SGD can only
accommodate a mini-batch of $32$ and $8$. Encouragingly, \reweight is
able to reduce the  memory bloat problem of DP-SGD by an average $3.8\times$
thanks to its
reweighted gradient derivation (\algo{algo:dpsgd}). 
This enables \reweight to achieve similar levels of maximum possible
mini-batch size that is feasible with non-private SGD.

\emph{Key takeaways: Training with DP-SGD as-is incurs an intractable amount of
	memory consumption, severely limiting the largest mini-batch size it can be trained on.
		\reweight can help achieve commensurate level of memory allocation to that
		of SGD, enabling much larger mini-batches to be employed for training.
		The memory efficiency of \reweight, however, comes at the cost of an additional pass of backpropagation for deriving per-batch weight gradients; for \reweight
		to become a practically viable solution for DP training, the
		added computation overhead of per-batch backpropagation should incur reasonable latency overheads.
}

\subsection{Identifying the Bottlenecks in  DP-SGD}
\label{sect:char_tpu_training_time_breakdown}

\begin{figure}[t!] \centering
\includegraphics[width=0.485\textwidth]{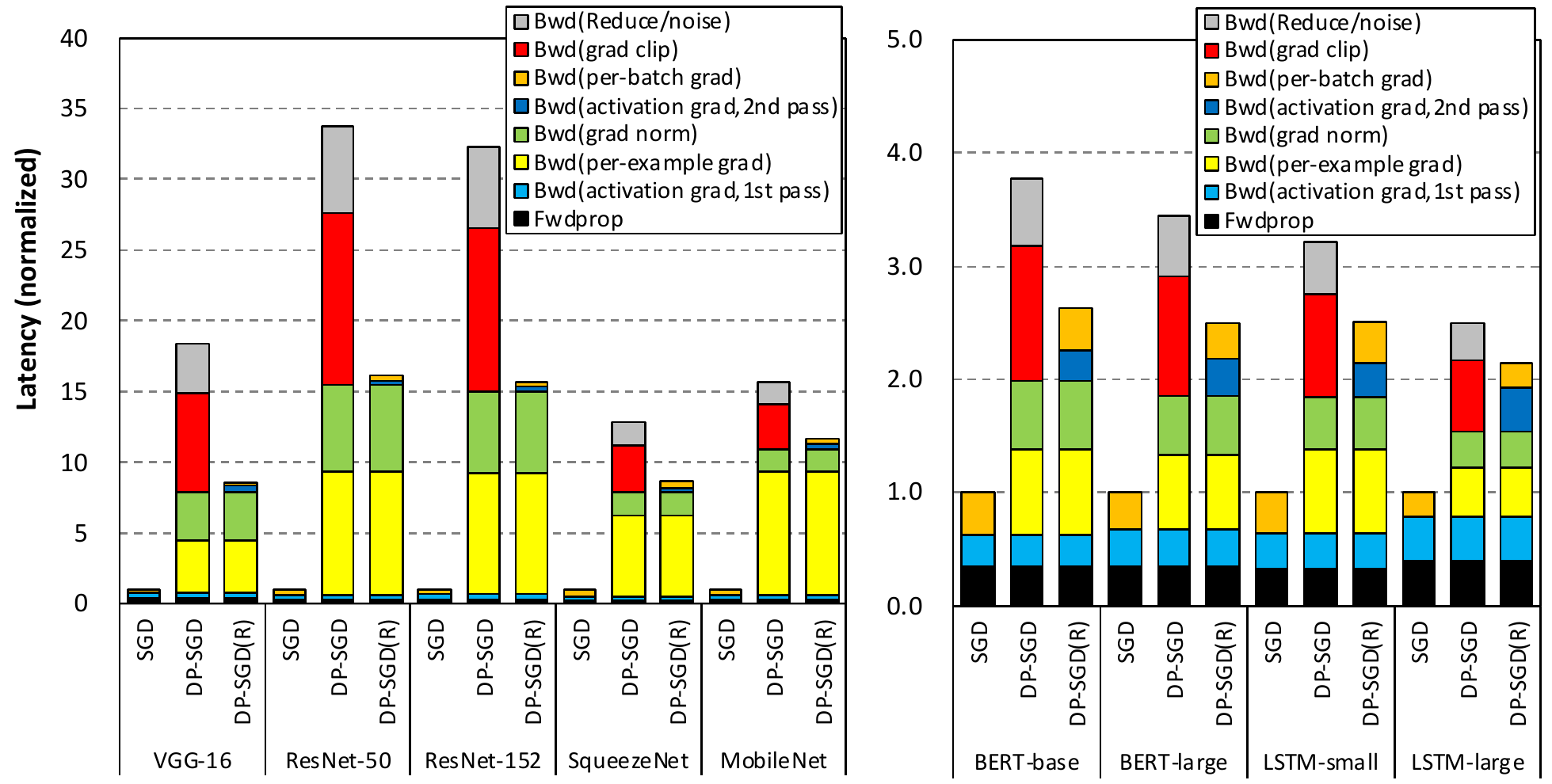}
\caption{Breakdown of SGD vs. DP-SGD training time into key stages of
	forward and backpropagation. The training time is normalized to non-private
SGD. Training mini-batch size is configured as the maximum mini-batch size possible
with DP-SGD
under Google TPUv3's HBM capacity (i.e., $16$ GB), which all three algorithms
employ identically. We utilize our cycle-level simulation framework modeled after
Google TPUv3 for this experiment.}
\vspace{-0.5em}
\label{fig:char_tpu_training_time_breakdown}
\end{figure}

In order to identify performance bottlenecks of DP-SGD,
	 \fig{fig:char_tpu_training_time_breakdown} explores the end-to-end training
	 time broken down into key steps of forward and backpropagation\footnote{
While we employ a simulation based methodology to breakdown  training time,
we confirmed that the key observations made in this subsection
	holds true in real Google TPUv3 experiments. Nonetheless, the black-box nature of Google TPUv3
	hardware/software system as well as the several caveats existing in its profiling tool (TensorBoard~\cite{tensorflow2015-whitepaper}) makes it challenging to precisely breakdown training time, so we report
	our results based on simulation.
	 }.  
We make	 several key observations from this characterization study.

\begin{figure*}[t!] \centering
\includegraphics[width=0.98\textwidth]{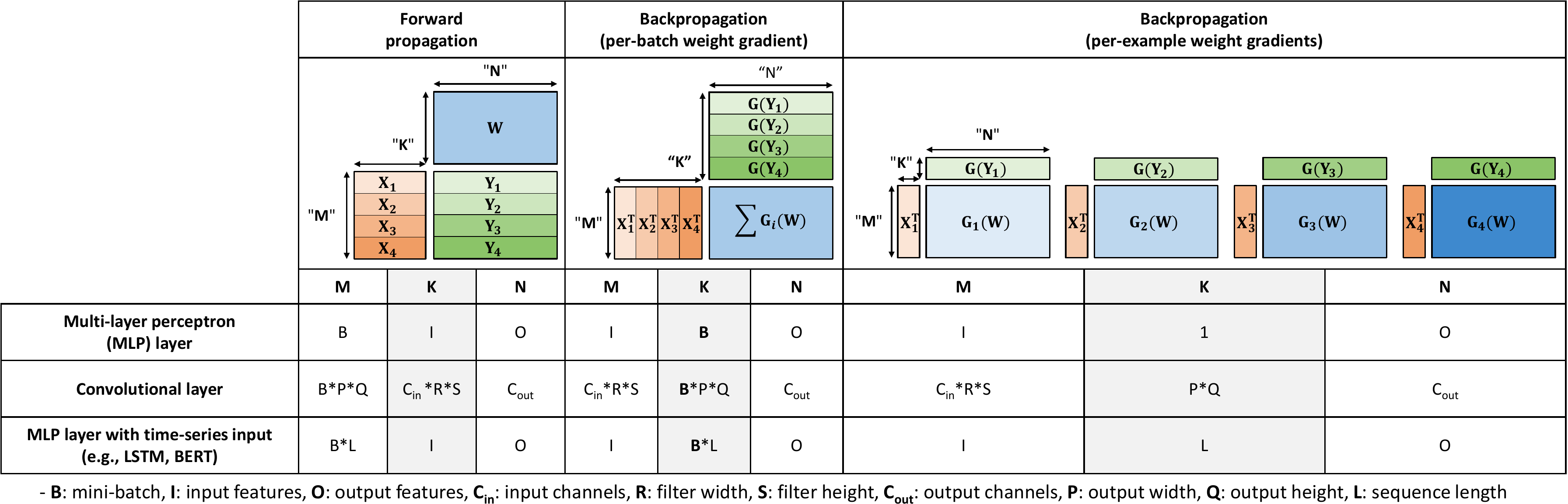}
\caption{Comparison of GEMM operation during (left) forward propagation and backpropagation for deriving (middle) per-batch weight gradients
	and (right) per-example weight gradients.}
\vspace{-0.5em}
\label{fig:overview_gemm_per_sample}
\end{figure*}

\begin{enumerate}

\item Both DP-SGD algorithms incur an average $9.1\times$/$5.8\times$
increase in training time vs. SGD, largely due to its significantly longer
backpropagation time (i.e., the forward propagation stages are practically identical
		among the three design points).  Unlike the non-private SGD where
backpropagation ``only'' accounts for $60-77\%$ of training latency, the
proportion of backpropagation takes up close to an average $99\%$ of latency under
DP-SGD, making it the single most important bottleneck.  

\item	 The significantly slower backpropagation of DP training is mostly
attributed to deriving the per-example weight gradients and
gradient post-processing.  Specifically, DP-SGD and
\reweight causes an average $12.7\times$ and $8.0\times$ increase in
latency for deriving the final weight gradient set \gw used for model updates.
It is worth pointing out that, regardless of which
				training algorithm is being employed, \emph{all} the gradients derived during
				backpropagation (i.e., \gx, \gw for both per-batch and per-example
					weight	gradients) are generated by conducting GEMMs (detailed further in
							\fig{fig:overview_gemm_per_sample}). 

\item Lastly, despite having to execute another backpropagation pass, the
reweighted \reweight surprisingly performs better than DP-SGD with an average
$31\%$ reduction in training time vs. DP-SGD. Reason for \reweight's
superior performance is as follows. The gradient post-processing stages of
DP-SGD involves a series of memory bandwidth limited operations, incurring high
latency.  \reweight, however, \emph{fuses} the gradient clipping/reduction stages of
post-processing as part of the ``reweighted''
	\emph{per-batch} weight gradient derivation, significantly reducing its
	memory traffic and reducing latency. In other words, while DP-SGD does not
	require \reweight's second backpropagation pass, the high latency overhead of
	gradient clipping/reduction stage (which is eliminated with \reweight,
			line $39$ in \algo{algo:dpsgd}) washes out DP-SGD's performance
	advantage vs. \reweight, rendering \reweight to perform superior than DP-SGD.

\end{enumerate}

\sethlcolor{pink}

\emph{ Key takeaways: DP-SGD incurs
	an order of magnitude higher training time than SGD because of the
		high latency incurred during the derivation of 1) per-example weight
		gradients and 2) gradient post-processing. And while both DP-SGD algorithms perform poorly vs.
		SGD, the reweighted \reweight demonstrated its superiority over DP-SGD as
		it not only provides consistently higher performance but it also helps drastically
		reduce its overall memory consumption, becoming a strong baseline DP
		training algorithm.  }

\begin{figure}[t!] \centering
\includegraphics[width=0.485\textwidth]{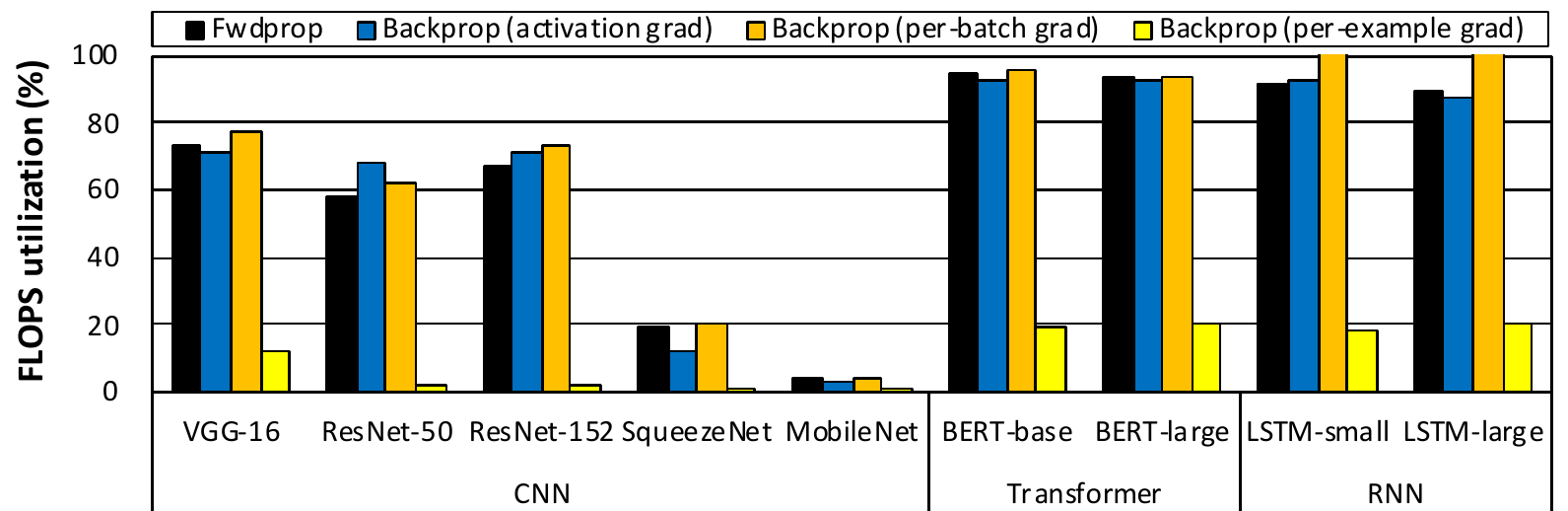}
\caption{
Google TPUv3's compute utilization during key GEMM operations of forward and backpropagation. We quantify TPU's compute utilization by
measuring the effective
	FLOPS  achieved vs. maximum available FLOPS in Google TPUv3.
}
\vspace{-0.5em}
\label{fig:char_tpu_compute_utilization}
\end{figure}

\subsection{Understanding the Bottlenecks in DP-SGD}
\label{sect:characterization_compute}

Our characterization in \sect{sect:char_tpu_training_time_breakdown}
identified two key bottlenecks of DP-SGD: 1) derivation of the per-example weight gradients
and 2) computing gradient norms. Below we root-cause the key reasons behind
such performance bottleneck.

{\bf Low compute utilization in DP-SGD backprop.} As mentioned in
\sect{sect:char_tpu_training_time_breakdown}, both per-batch and per-example
weight gradients are derived using GEMMs. To understand the reason behind
GEMM's low throughput during DP-SGD backpropagation, we compare the GEMM
dimensions of forward and backpropagation for both SGD and DP-SGD in
\fig{fig:overview_gemm_per_sample}\footnote{It is worth pointing out that
	\reweight, our highest performing DP-SGD design point, utilizes \emph{both}
	per-example (during the 1st backpropagation phase) and per-batch (during the
			2nd phase) weight gradient derivations for computing the final \gw, all
		of which are covered by the GEMMs illustrated in
		\fig{fig:overview_gemm_per_sample}.}. As depicted, deriving a per-batch
		weight gradient is equivalent to a GEMM with its K-dimension scaling
		proportional to the mini-batch size, generating a \emph{single} set of \gw
		(i.e., the inner-product along the K-dimension has the effect of
		 conducting a gradient reduction across all mini-batch examples).
		Consequently,  the systolic-array is provided with more abundant data-level
		parallelism and weight reuse opportunity under larger mini-batched GEMMs,
		better saturating its compute units.  Contrast that with the GEMMs that
		derive the per-example weight gradients where a total of $B$(=mini-batch)
	independent GEMMs are conducted (generating $B$ sets of \gw), each GEMM's
	K-dimension sized irrespective of the mini-batch size $B$ exhibiting irregularly
	shaped GEMMs.

	Now, recall from \sect{sect:background_systolic} that GEMMs with small
	K-dimensions map poorly to the highly regular and structured design of
	systolic arrays, leading to significant underutilizations (\fig{fig:systolic_dataflow}). In
	\fig{fig:char_tpu_compute_utilization}, we quantify the magnitude of systolic
	array's underutilization across all major GEMM operations of forward and
	backpropagation. Across all studied DNN models, the irregularly shaped GEMM
	for per-example weight gradients consistently exhibit much lower compute
	utilization compared to the other GEMM operations (i.e., forward propagation
			as well as backpropagation for deriving the input activation gradient and per-batch
			weight gradient).  These results explain why the per-example weight
	gradient derivation incurs such high performance overhead
	(\fig{fig:char_tpu_training_time_breakdown}), providing important guidelines
	on designing an accelerator tailored for the unique dataflow of DP-SGD. 

{\bf Memory-bound gradient norm derivation.} The post-processing stages of
DP-SGD (i.e., gradient norm derivation, gradient clipping/reduction) are all
memory-bound operations with low compute intensity. With reweighted \reweight
established as our baseline DP-SGD algorithm, computing the gradient norms is
the only major step left in our memory-bound bottleneck stage during gradient
post-processing (see \fig{fig:char_tpu_training_time_breakdown}).  Aside from
the systolic array engine, Google TPUv3 comes with an on-chip vector processing
unit that handles vector operations (e.g., vector additions and multiplications),
which Google TPUv3 utilizes for computing
	gradient norms.  We observe that the per-example weight gradients that are
	targeted for gradient norm derivation are typically too large to be stored
	inside Google TPUv3's on-chip buffers, rendering these per-example gradient tensors to be
	spilled to off-chip DRAM.  This leads to frequent off-chip memory accesses
	to fetch the weight gradients for computing gradient norms, exhibiting memory
	bandwidth limited behavior and causing steep latency penalties.

\emph{Key takeaways: The two key bottleneck operations of DP-SGD are 1)
	series of irregular, tall-skinny GEMMs with small K-dimension sizes during
		per-example weight gradient derivation, and 2)
		 memory-bandwidth limited vector reduction operations for deriving gradient norms.  } 

\section{DiVa Architecture and Design}
\label{sect:diva_arch}

\begin{figure}[t!] \centering
\includegraphics[width=0.485\textwidth]{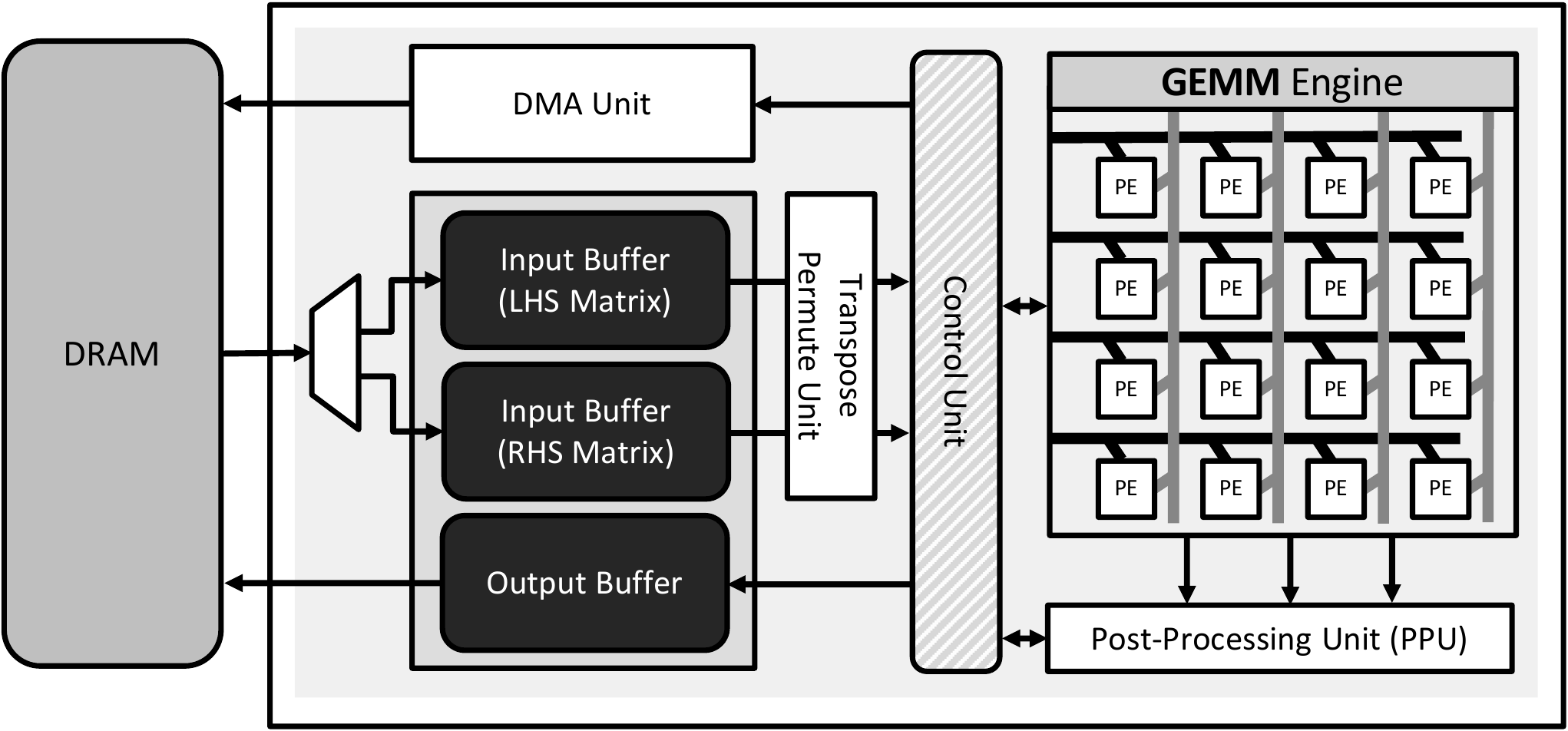}
\caption{
High-level overview of \proposed.
}
\vspace{-0.5em}
\label{fig:proposed_diva_overview}
\end{figure}

\subsection{Architecture Overview}
\label{sect:proposed_overview}

\proposed consists of a GEMM engine which is implemented as a 2D spatial array
of PEs, a post-processing unit (PPU) for accelerating gradient post-processing,
	 a large on-chip SRAM buffer that is partitioned for storing the left-hand
	 side (LHS) and right-hand side (RHS) input matrices as well as one output
	 matrix, a transpose/permute unit that handles matrix transposition or im2col
	 operation (i.e., transforms convolutional layers into
			 GEMM)~\cite{chetlur2014cudnn,jia2014learning},
	 a DMA unit that orchestrates on-/off-chip data movements, and the main
	 control unit (\fig{fig:proposed_diva_overview}).  The control unit populates
	 the on-chip SRAM buffer as appropriate per \proposed's tiled GEMM execution
	 order. Once the two input matrix tiles are uploaded, the control unit
	 initiates the outer-product dataflow based matrix multiplication using the
	 GEMM engine.  After the GEMM computation is finished, depending on which
	 phase of DP training \proposed is currently under processing, either the
	 output activation (forward propagation), the per-batch input gradient, the
	 per-batch weight gradient, or the per-example weight gradient
	 (backpropagation) is derived and latched inside the spatial PE array, i.e., \proposed is classified as an \emph{output stationary} (OS) dataflow. In
	 cases where gradient post-processing over per-example weight gradients is
	 required, the control unit directly routes the outputs latched inside the
	 PEs into the PPU. The PPU output is then drained back into the
	 SRAM buffer and finally the off-chip DRAM by the DMA unit.

\subsection{Outer-product GEMM Engine}
\label{sect:proposed_dataflow}

{\bf Challenges of systolic array dataflow.} A fundamental limitation of the
systolic array architecture is that it suffers from very low PE utilization
when the K-dimension of the LHS (and RHS) matrix is small, which is a unique
property of the GEMMs that derive DP-SGD's per-example weight gradients. Under
the WS systolic dataflow, having a small K-dimension results in only partially
latching the RHS matrix rows within the systolic array and for the OS systolic
dataflow, small K-dimension leads to a shorter length (LHS/RHS) vectors being
streamed into the OS systolic array (\fig{fig:systolic_dataflow}). This leads to
only a handful of PEs being utilized for MAC operations each cycle, far lower
than the maximum MAC throughput available across the systolic array, thereby
significantly underutilizing its compute power.

\begin{figure}[t!] \centering
    \subfloat[]
    {
\includegraphics[width=0.485\textwidth]{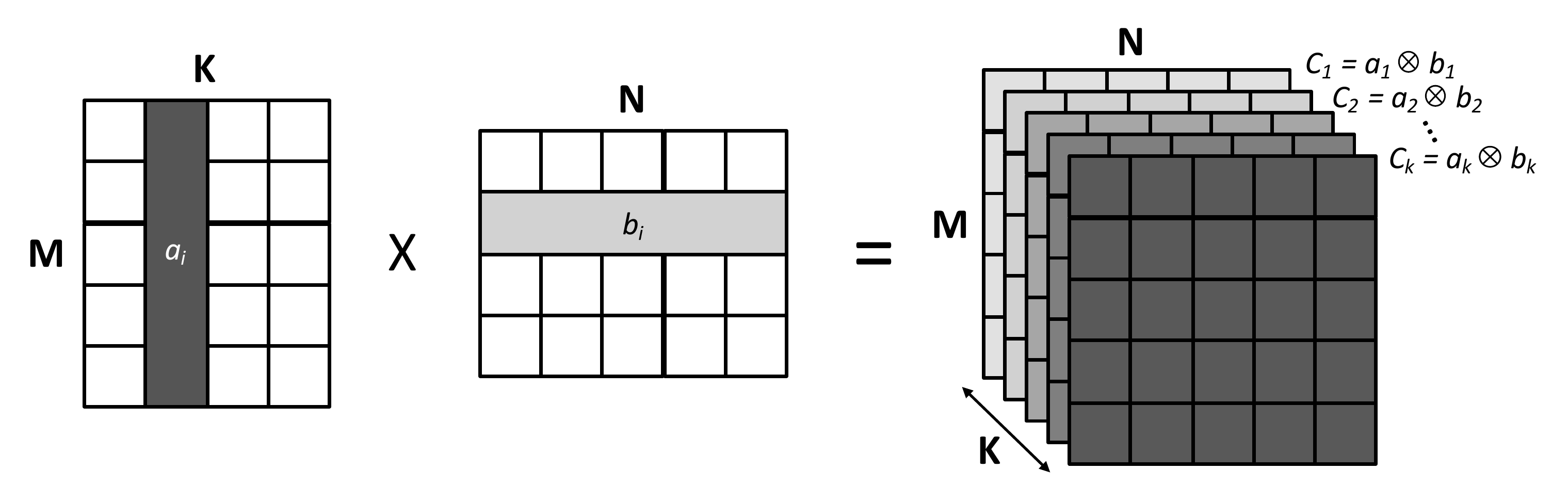}
    }
    \vspace{-0em}
    \subfloat[]
    {
\includegraphics[width=0.41\textwidth]{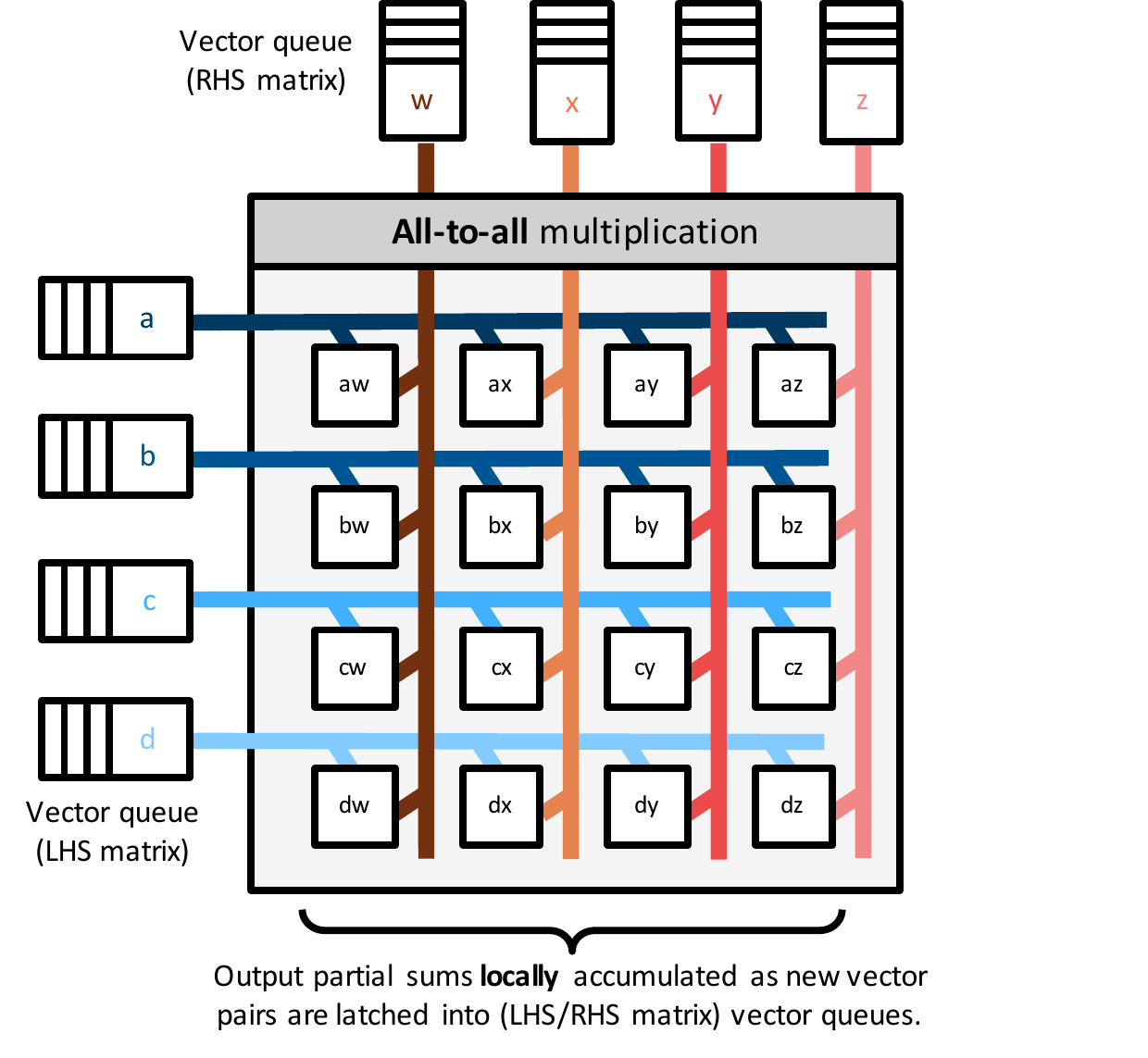}
    }
    \vspace{-0em}
\caption{(a) An outer-product matrix multiplication between (M,K) matrix $A$ and (K,N) matrix $B$. (b) Implementation of an
	outer-product based GEMM engine via 2D spatial array of PEs conducting all-to-all multiplication.
		}
    \label{fig:proposed_outer_product}
		\vspace{-0.5em}
\end{figure}

{\bf Outer-product dataflow and its implementation.} To address such challenge, \proposed proposes a GEMM engine based on the
\emph{outer-product} dataflow.  \fig{fig:proposed_outer_product}(a) shows an 
outer-product GEMM between matrices $A$ and $B$, where the matrix
multiplication is decomposed into series of outer-product multiplications between pairs
of vectors, $a_{i}$ and $b_{i}$. Specifically, each column of $A$ ($a_{i}$) and the
corresponding row of $B$ ($b_{i}$) is multiplied in an all-to-all manner, producing $K$
partial sum matrices, $C_{i}$. These partial sum matrices are summed altogether
to produce the final output matrix $C$.  Notice how the outer-product 
generates a total of $M$$\times$$N$ MAC operations over two input vectors with
lengths $M$ and $N$, respectively. Unlike an inner-product where two input vectors
		 must be of equal length, outer-product can have arbitrary length
		 input vectors $M$ and $N$, having more flexibility and robustness in mapping the GEMM across
		 the PE array.

\proposed seeks to address the PE underutilization issue of systolic arrays
by leveraging the aforementioned property of outer-product dataflow as
illustrated in \fig{fig:proposed_outer_product}(b). For clarity of explanation,
						suppose the M-/N-dimension sizes of an (M,K,N) GEMM matches the
						spatial PE array's height (PE$_{H}$) and width (PE$_{W}$).  The
						frontend of \proposed's GEMM engine streams in two vectors of
						length $M$ (columns of LHS matrix) and $N$ (rows of RHS matrix) each
						clock cycle. Each row and each column of our spatial PE array
						utilizes its local bus to broadcast the streamed in vectors across
						the PEs, conducting an \emph{all-to-all} multiplication as depicted
						in \fig{fig:proposed_outer_product}(b).  The partial sum matrix
						$C_{i}$ (i.e., matrix $M$$\times$$N$) gets newly generated every
						clock cycle, which are accumulated \emph{locally} within each
						individual PE, requiring a total of $K$ clock cycles to derive the
						final matrix $C$ for the (M,K,N) dimension GEMM.  Note how
						\proposed's GEMM engine is \emph{always} capable of conducting
						$M$$\times$$N$ MAC operations each cycle, \emph{regardless of the K-dimension size}. Again, it is worth pointing out that the outer-product
						dataflow broadly falls under an OS dataflow
						as the final output remain stationary within the spatial PE array during 
						GEMM computation.

Given DP-SGD's key bottleneck is the small K-dimension GEMMs in deriving
per-example weight gradients, \proposed's outer-product GEMM engine can
significantly improve performance as the effective throughput of these GEMMs
are dependent upon the M-/N-dimensions, and not the K-dimension (see
		\fig{fig:overview_gemm_per_sample}). In \sect{sect:eval_perf}, we quantify the
magnitude of how much improvement \proposed brings about in closing the wide
performance gap between SGD vs.  DP-GSD.

\begin{figure}[t!] \centering
\includegraphics[width=0.44\textwidth]{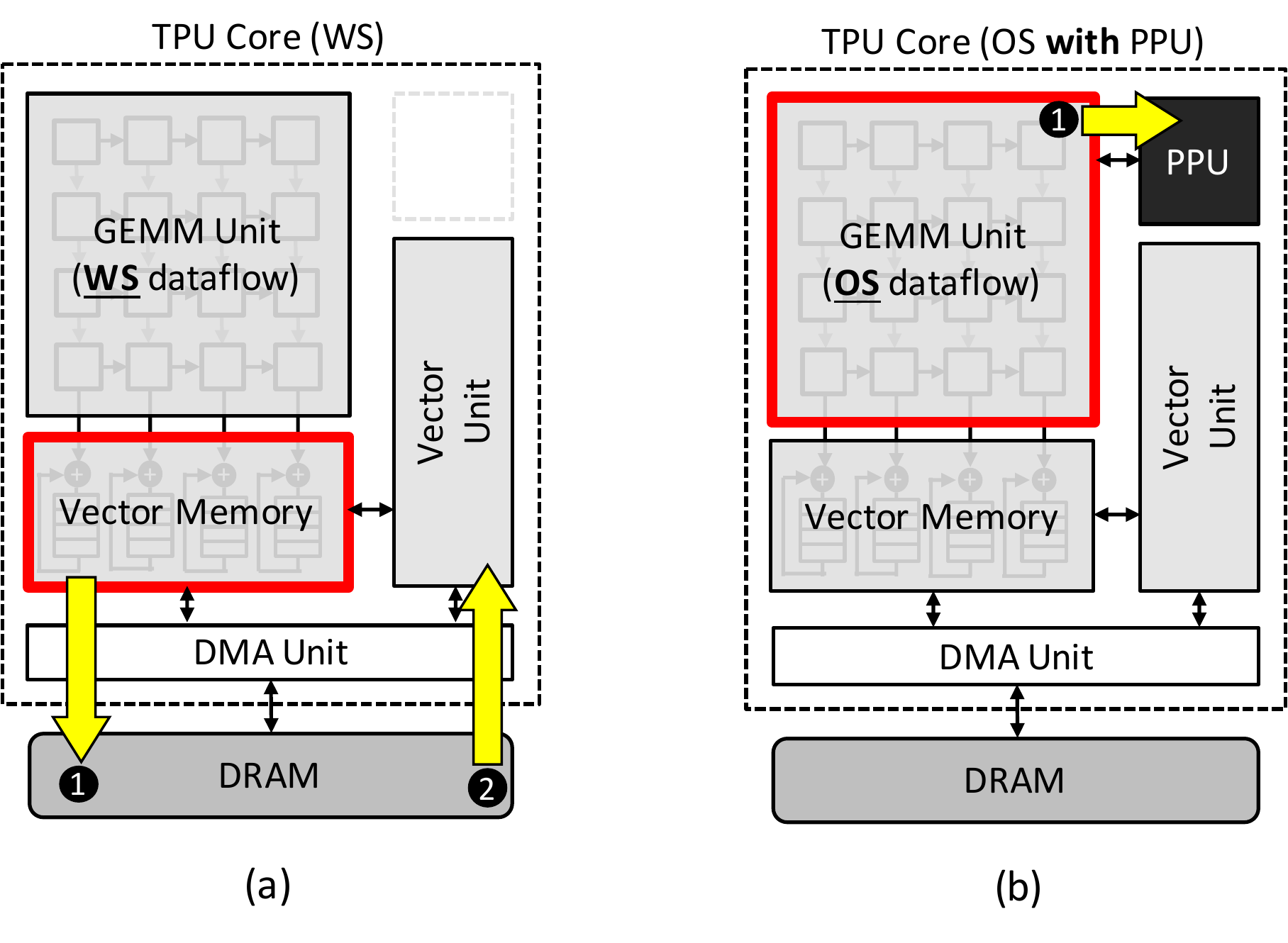}
\vspace{0em}
\caption{
Dataflow of gradient norm derivation under (a) a WS systolic array and (b) an OS systolic array. 
	The red boxes indicate locations where the per-example weight gradients are stored.
}
\vspace{-0.5em}
\label{fig:motivation_ppu}
\end{figure}

\subsection{Post-Processing Unit (PPU) Design}
\label{sect:proposed_ppu}

{\bf WS vs. OS dataflow in gradient norm derivation.} Another crucial
bottleneck in DP-SGD training is the memory-bandwidth limited gradient norm
derivation (\sect{sect:characterization_compute}). Before discussing
\proposed's PPU, let us first discuss the challenges of
Google TPU's WS dataflow in handling gradient norm computation, root-causing
its memory-bound characteristic.  As discussed in
\fig{fig:systolic_dataflow}(c), for WS systolic arrays to achieve high PE
utility, the length of the LHS matrix streamed in from the left side of the
GEMM engine must be sufficiently large enough to amortize the effect of idle
cycles manifested in the diagonal direction of the input stream.   An important
implication of having large LHS input streams is that the output SRAM buffer
that temporarily stores the WS systolic array's output must be sufficiently
large, proportional to the size of the LHS input stream. In Google TPUv3, the
size of this SRAM buffer (referred to as \emph{Vector Memory} in
		TPUs~\cite{norrie2020google,jouppi2020domain}, \fig{fig:motivation_ppu}) is $16$ MB, accounting for the largest on-chip memory capacity.  Since the
per-example weight gradients, subject for gradient norm derivation, are stored
inside this (large) SRAM buffer, the control unit can either 1) directly
forward this several tens of MBs worth of tensors to the vector unit for
on-the-fly gradient norm derivation, or 2) temporarily spill them to off-chip
DRAM and process them later on. Careful examination of  this process over
Cloud TPUs revealed
that Google TPUv3 typically takes the latter approach and spills these large
sized tensors to DRAM (step $1$ in \fig{fig:motivation_ppu}(a)), later fetching
them back on-chip for gradient norm derivation (step $2$). We speculate the
reason for such design decision is as follows. Directly forwarding the
per-example gradients to the vector unit for on-the-fly gradient norm
derivation requires the GEMM engine (systolic array) to remain stalled until
the gradient norm computation is finalized by the vector unit (i.e., unless the output SRAM buffer is
		vacant, the systolic array does not have a temporary buffer to store
		the \emph{next} GEMM's output). Since these tensors are sized
		in the tens of MBs scale, deriving gradient norms on-the-fly itself incurs
		high latency, which directly affects the main GEMM unit's stalled period.
Double-buffering the output SRAM buffer, however, is practically a non-option since
this buffer is already in the order of tens of MBs, due to the nature of a WS dataflow.

\begin{figure}[t!] \centering
\includegraphics[width=0.46\textwidth]{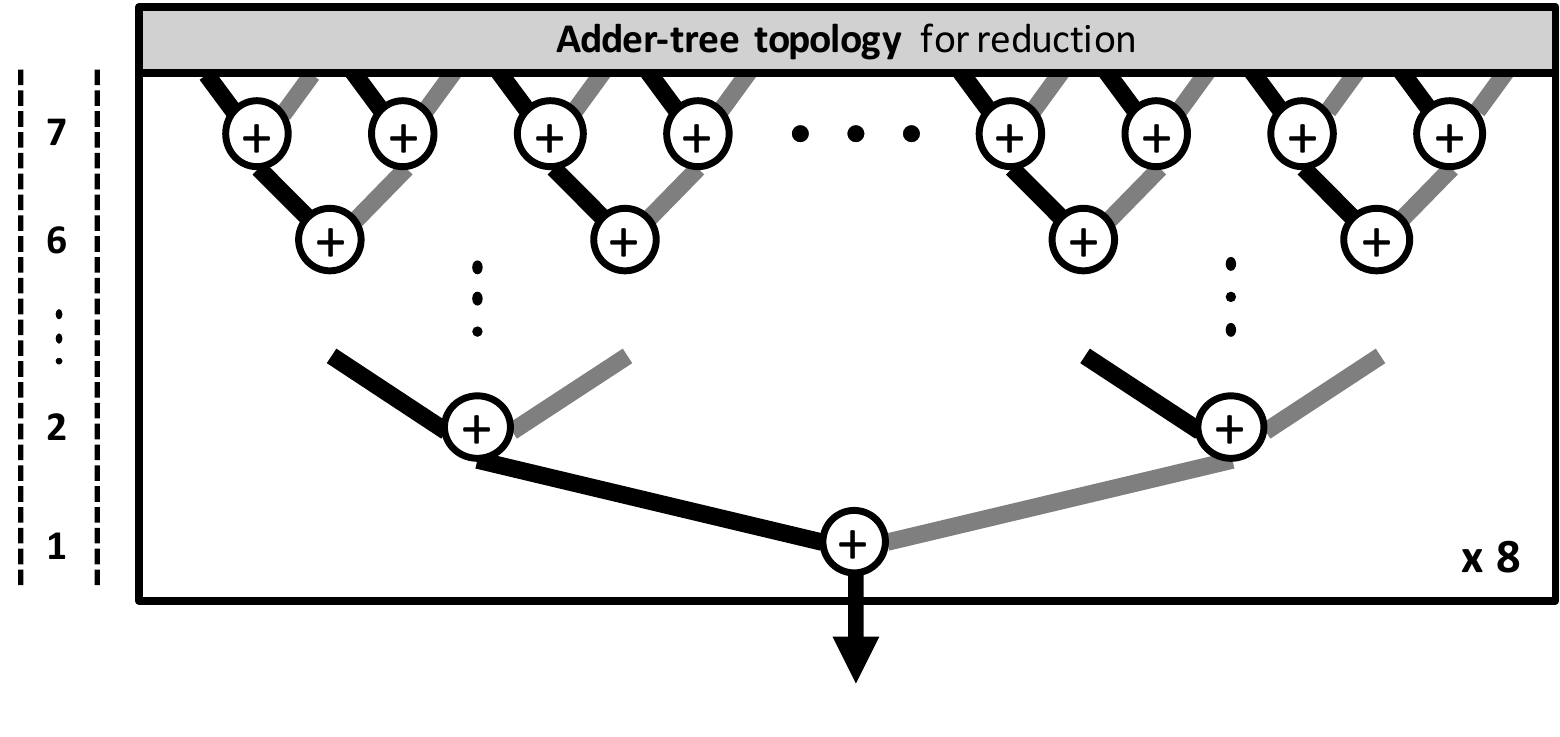}
\vspace{-0em}
\caption{
\proposed's multi-level ($=7$) adder-tree in PPU, assuming ($128$,$128$) PE array (i.e., $2^{7}=128$). The baseline \proposed configuration employs a PPU that can read out $8$ output rows from the GEMM engine, so $8$ instances of adder-trees are instantiated within a PPU.
}
\vspace{-0.5em}
\label{fig:proposed_adder_tree}
\end{figure}

Consequently, an OS systolic dataflow becomes an appealing alternative for handling
on-the-fly gradient norm derivation. As illustrated in
\fig{fig:motivation_ppu}(b), the per-example weight gradients are derived in a
much smaller, finer-granularity in an OS dataflow, the size of which scales
proportional to the systolic array size. Under the ($128$,$128$) PE array, this
amounts to ``only'' $64$ KB (=$128\times128\times4$ Bytes), far less than
the tens of MBs of tensors requiring post-processing under a WS dataflow. Overall,
		the benefits of an OS dataflow that can directly forward the per-example
		weight gradients to the vector unit is clear: 1) the tensors no longer have
		to be spilled/retrieved to/from DRAM, alleviating its memory bandwidth
		pressure, and 2) the datapath of such on-the-fly derivation of gradient
		norm opens up an opportunity to further boost its overall throughput with a
		dedicated accelerator microarchitecture tuned for gradient norm derivation.
We now detail \proposed's PPU design, readily applicable not only for \proposed's
outer-product GEMM engine (i.e., outer-product also falls under an OS dataflow) 
	but also for an OS dataflow based systolic arrays.

{\bf PPU architecture.} Deriving an L2 norm of a tensor (\eqn{eqn:norm})
	requires an element-wise multiplication of the target tensor with itself,
	followed by a reduction operation over all of its elements to generate a
	single output scalar value.

\begin{equation}
\Vert g \Vert_2 = \sqrt{\sum\limits_{i_1=1}^{d_1} \cdots \sum\limits_{i_n=1}^{d_n} g_{i_1 \ldots i_n}^2}
\textrm{ ,\quad where $g \in \mathbb{R}^{d_1 \times \cdots \times d_n}$}
\label{eqn:norm}
\end{equation}

While conventional vector units do an excellent job in the element-wise
dot-product operation, they become sub-optimal in conducting reductions as it
require multiple iterations of vector permutations to retrofit reductions as
vector operations.  In \proposed's PPU, we implement a spatial, multi-level
adder-tree based reduction unit for accelerating gradient norm derivation as
illustrated in \fig{fig:proposed_adder_tree}.  Under such tree-based
topological design, the input data loading and output data generation time is
in the order of O($1$) and O(log$_{2}$$E$), respectively ($E$: number of elements to
		reduce), significantly reducing latency for gradient norm derivation.

\sethlcolor{yellow}
\begin{figure}[t!] \centering
\includegraphics[width=0.41\textwidth]{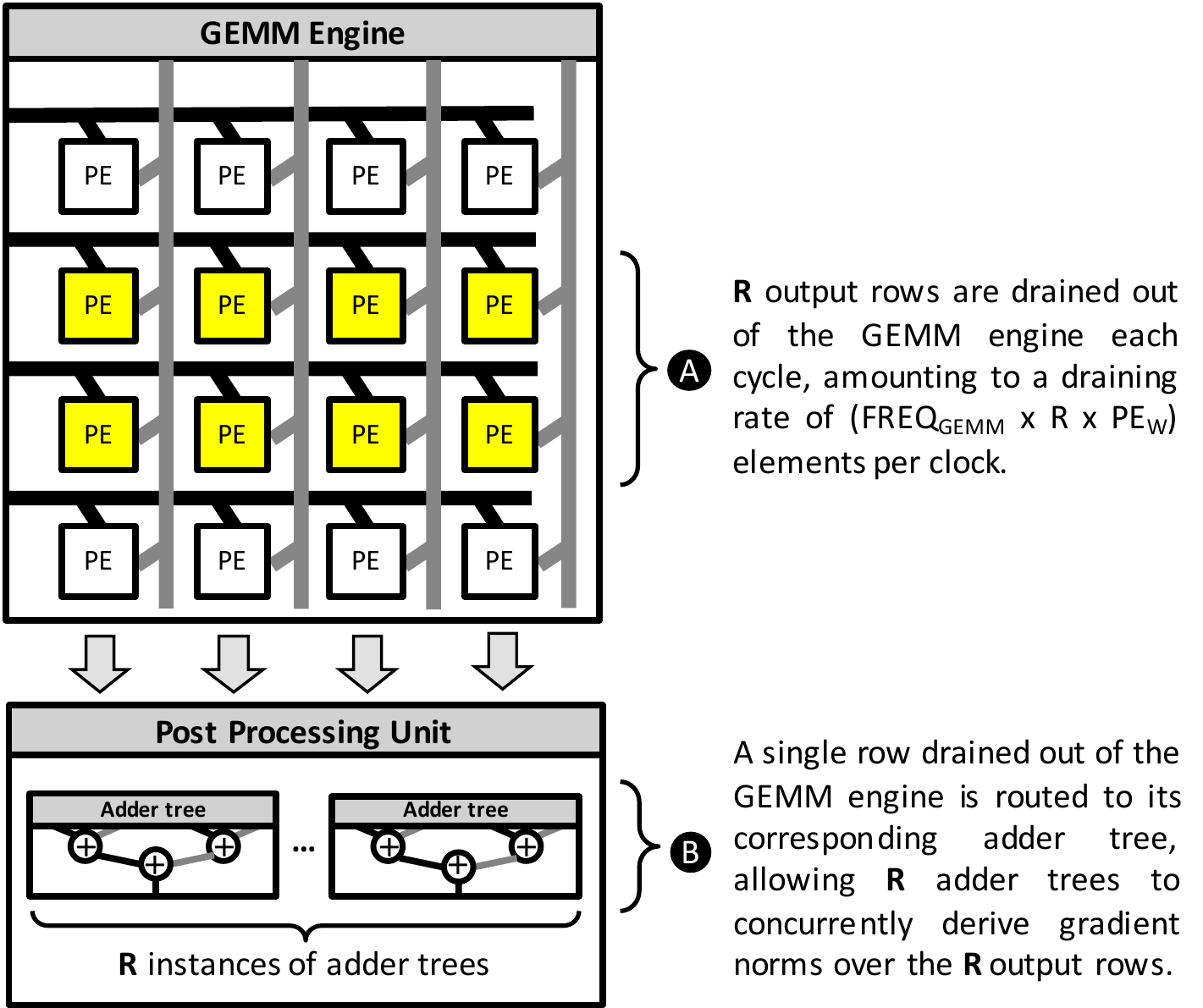}
\vspace{0.5em}
\caption{
Interfacing DiVa's GEMM engine and PPU for seamless gradient norm derivation.
	Under default DiVa configuration, both units are synthesized
		at the same operating frequency ($=940$ MHz) under $65$ nm technology.
}
\vspace{-0.5em}
\label{fig:revision_rA_Q1}
\end{figure}

\sethlcolor{yellow}
{\bf Interface between GEMM engine and PPU.} \fig{fig:revision_rA_Q1}
summarizes 1) the rate in which DiVa's GEMM engine drains out per-example
	weight gradient vectors to the PPU, and 2) the required PPU's processing
		throughput to seamlessly derive gradient norms.  As depicted, DiVa's GEMM
		engine reads out $R$ output rows each clock cycle  and forwards them to the
		PPU for post-processing.  Assuming the GEMM engine's operating frequency is
		\texttt{FREQ$_{GEMM}$},
		(\texttt{FREQ$_{GEMM}$}$\times$\texttt{R}$\times$\texttt{PE$_{W}$}) elements
			are drained out of the GEMM engine each clock cycle.  Under DiVa's
			default configuration of \texttt{FREQ$_{GEMM}$}=$940$ MHz, $R$=$8$ rows,
		\texttt{PE$_{W}$}$=128$ elements/row, and $4$ Bytes/element, this amounts to
			($940$M$\times8\times128\times4$B)$=$$(3.85)$ TB/sec of weight
			gradients to reduce. DiVa's PPU is provisioned with sufficient processing throughput
			by having the reduction unit be designed as a
			$7$-level (=log$_2$\texttt{PE$_{W}$}), pipelined adder-tree with $R$
			separate instances of it incorporated within the PPU. Specifically,
		each output row of the GEMM engine is forwarded to its corresponding adder-tree 
			each clock cycle.
			Because
the operating frequency of PPU (\texttt{FREQ$_{PPU}$}) matches \texttt{FREQ$_{GEMM}$} 
	and
			the PPU's adder tree is capable of reducing 
$PE\_W$$=$$128$ elements each clock in a pipelined manner,
DiVa is able to seamlessly 
derive gradient norms. 
			Overall, a total of $128$/$R$ clock cycles are required in fully draining
			out the GEMM engine's outputs for gradient norm derivation.

\subsection{Design Overhead}
\label{sect:proposed_overhead}

While \proposed's outer-product engine provides robust
GEMM performance across a wide range of GEMM shapes, it can incur 
design overheads vs. systolic arrays in terms of 1) inter-PE communication
channels and 2) read/write bandwidth from/to on-chip SRAM buffers. 

One of the key advantages of systolic arrays is its simple inter-PE
communication datapath as only spatially nearby PEs exchange data amongst them,
							simplifying its design. \proposed's all-to-all multiplication
							requires each and every rows and columns to have a local bus
							datapath to broadcast the incoming two input vectors across the
							PEs, potentially incurring higher area and power overheads vs.
							systolic arrays. 

In terms of on-chip SRAM bandwidth needs, the WS systolic dataflow requires
sufficient SRAM read bandwidth to be provisioned for a one-time latching of the RHS matrix into
the systolic array (e.g., Google TPUv3 is capable of filling in $8$ rows/cycle, \tab{tab:comparison_sram_bandwidth}), accompanied by a single vector streaming bandwidth of
O(PE$_{H}$) to feed in the LHS matrix.  In contrast, \proposed outer-product
dataflow needs to stream in two separate vectors of length PE$_{H}$ and
PE$_{W}$ consistently to the GEMM engine, having  an O(PE$_{H}$+PE$_{W}$) SRAM
read bandwidth at steady state. It is worth pointing out that the
O(PE$_{H}$+PE$_{W}$) SRAM read bandwidth of outer-product (as well as its SRAM write
		bandwidth) is no worse than the OS systolic dataflow, as summarized in
\tab{tab:comparison_sram_bandwidth}   (see \fig{fig:systolic_dataflow}(c) vs.
		\fig{fig:proposed_outer_product}(b)). As we uncovered in the previous
subsection, the OS dataflow provides better opportunities than WS for
accelerating gradient norm derivation (but the OS in itself does not necessarily 
help accelerate	the GEMMs with small K-dimensions for
		per-example weight gradient derivation), rendering \proposed's higher
on-chip SRAM bandwidth vs. WS a reasonable trade-off.  We quantitatively
analyze the design overheads of \proposed in \sect{sect:eval_overhead},
				demonstrating its merits.

\begin{table}[tpb]
\centering
\caption{Comparison of on-chip SRAM buffer read/write bandwidth requirements assuming Google TPUv3 level configuration~\cite{norrie2020google,jouppi2020domain}.
Note that the LHS/RHS matrices are stored as $16$-bit data types ($2$-Bytes) while the accumulation happens in $32$-bits ($4$-Bytes) for 
	output derivation~\cite{kalamkar2019study}.
}
\vspace*{-0em}
\scriptsize
\begin{tabular}{|c|c|c|}
\hline
\multirow{2}{*}{\bf{Data type}} & \multicolumn{2}{c|}{\bf{Dataflow ({\bf B}ytes/clock)}}\\
\cline{2-3}
 & Systolic WS & Systolic OS \& Outer-product\\
\hline
\hline
Input LHS & PE$_{H}$$\times$2B & PE$_{H}$$\times$2B \\
Input RHS & PE$_{W}$$\times$8$\times$2B & PE$_{W}$$\times$2B \\
Output & PE$_{W}$$\times$4B & PE$_{W}$$\times$8$\times$4B \\
\hline
\hline
Total
& (2$\times$PE$_{H}$+20$\times$PE$_{W}$)B
& (2$\times$PE$_{H}$+34$\times$PE$_{W}$)B
\\
\hline
\end{tabular}
\vspace{-0.5em}
\label{tab:comparison_sram_bandwidth}
\end{table}

\section{Methodology}
\label{sect:methodology}

\sethlcolor{orange}
\begin{figure*}[t!] \centering
\includegraphics[width=0.995\textwidth]{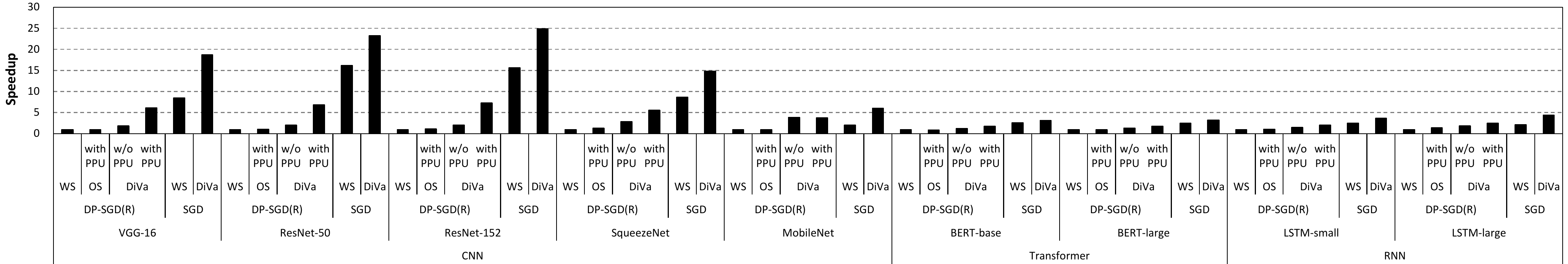}
\caption{
End-to-end speedup vs. baseline WS systolic array. DiVa is evaluated with/without PPU. Unlike WS, the OS systolic array can reap the benefits of PPU (\sect{sect:proposed_ppu}), so we evaluate OS with PPU implemented.
	We also present non-private SGD trained with WS and DiVa as a comparison point.
}
\vspace{-0.5em}
\label{fig:eval_speedup}
\end{figure*}

\sethlcolor{pink}

The workload characterization in \sect{sect:characterization} is
performed over the Google Cloud TPUv3 platform. We used TensorFlow Privacy
(v0.5.1)~\cite{tfprivacy} to compare SGD vs. DP-SGD's memory
allocation size and maximum possible mini-batch size.
When measuring performance,
we construct a strong baseline configuration that represents state-of-the-art
by employing JAX (v0.3.7)~\cite{jax2018github} with auto-vectorization features
enabled~\cite{SUB_2021,KUR_2022}, allowing DP-SGD to suffer less
from the small K-dimension GEMMs during per-example weight gradient derivation.
The JAX auto-vectorization enhanced GEMM kernels are utilized when measuring
all baseline systolic WS/OS as well as our baseline GPU systems.
Driven by our
characterization, we employ the following measures for estimating performance,
	area, and energy in \sect{sect:results}.

{\bf Performance.} We developed a cycle-level simulator for both \proposed and Google TPUv3
as described in \cite{norrie2020google,jouppi2020domain,tpu_patent1,tpu_patent2,tpu_patent3,tpu_patent4}. The
TPUv3 performance model is carefully validated against Google Cloud TPUv3 in terms
of effective throughput across a wide range of GEMM shapes (Pearson
		correlation coefficient: $0.95$). \tab{tab:diva_config} summarizes the
key parameters of \proposed's baseline architecture design, which is configured similar to
Google TPUv3.

{\bf Area/power.} To estimate the area and power of \proposed, we implement both the WS and OS systolic
array along with our outer-product GEMM engine augmented with the PPU in RTL using SystemVerilog.
The RTL is synthesized with Synopsys
Design Compiler targeting $0.94$ GHz of operating frequency using a $65$ nm
standard-cell library. 

{\bf Energy.} The energy consumed in \proposed's GEMM engine and PPU is
estimated by the power numbers from Synopsys Design Compiler.
For modeling the power and energy consumption of on-chip SRAM usage, we use CACTI~\cite{cacti} targeting
a $65$nm process. We  
employ the energy model from Horowitz~\cite{horowitz:isscc}
for quantifying energy per operation for off-chip DRAM accesses.

{\bf Benchmarks.}
We study five DNN models for computer vision (VGG,
		ResNet-50, ResNet-152, SqueezeNet, MobileNet)~\cite{vggnet,resnet,squeezenet,mobilenet} and 
four models for natural language processing (BERT-base/large, LSTM-small/large)~\cite{bert,lstm,charlstmclassification}.
State-of-the-art DP-SGD algorithms for computer vision are currently demonstrated with
its efficacy over CIFAR-10 datasets so the evaluation in \sect{sect:results} assumes such
setting. We discuss \proposed's efficacy when deviating
from our default configuration in \sect{sect:eval_sensitivity}.

\begin{table}[t!]
  \centering
  \caption{\proposed architecture configuration.}
\scriptsize
  \begin{tabular}{|c|c|}
		\hline
		\multicolumn{2}{|c|}{\textbf{Processor architecture}} \\
		\hline
    PE array dimension     			& $128 \times 128$   \\
    \hline              
    PE operating frequency				& $940$ MHz \\
    \hline              
    On-chip SRAM size   			& $16$ MB \\
    \hline              

    \multicolumn{2}{|c|}{\textbf{Memory subsystem}} \\
    \hline
    Number of memory channels 	& $16$	\\
    \hline
    Memory bandwidth	& $450$ GB/sec	\\
    \hline
    Memory access latency & $100$ cycles  \\
		\hline
  \end{tabular}
\vspace{-.5em}
  \label{tab:diva_config}
\end{table}

\section{Evaluation} 
\label{sect:results}

\sethlcolor{yellow}
As we demonstrated DP-SGD(R)'s superiority over a vanilla DP-SGD, 
	 this section always employs DP-SGD(R) as the baseline differentially private training algorithm.

\sethlcolor{orange}
\subsection{Performance}
\label{sect:eval_perf}

{\bf End-to-end speedup.} \fig{fig:eval_speedup} summarizes the end-to-end
speedup offered with \proposed. In general, \proposed exhibits consistently
higher performance than both WS and OS systolic array, achieving an average
$3.6\times$ (max $7.3\times$) speedup over WS. Such high speedup enables
privacy-enhanced \proposed to reach an average $75\%$ of the performance
of a non-private SGD trained with systolic WS, closing their wide performance gap
(\fig{fig:char_memory_usage}). In fact, for DNNs that baseline WS 
especially suffers from low performance (MobileNet, LSTM-large), \proposed's
DP training actually performs better than non-private SGD.  Note
that \proposed \emph{without} PPU, while still providing meaningful speedup,
		 leaves significant performance left on the table (e.g., \proposed
				 with/without PPU achieves $7.3\times$/$2.1\times$ speedup vs.
				 WS in ResNet-152), demonstrating the importance of optimizing gradient
		 post-processing.

It is also interesting to note that DiVa utilized for training ``non-private SGD''
	(denoted \texttt{DiVa-SGD}) performs better than systolic WS used for non-private
	SGD training, achieving an average $1.6\times$ higher performance. Such superior 
	performance comes from DiVa's outer-product dataflow, which enables our architectural
	substrate to be robust for small K-dimension GEMMs existent in non-private SGDs.
	As we later discuss in \sect{sect:eval_overhead}, however, 
	DiVa's outer-product does come at a higher area overhead than the baseline systolic WS.
	So from a performance/area perspective, ML accelerators optimized for non-private SGD training only
	might	prefer the lightweight
		systolic design than our proposed outer-product dataflow.
		Nonetheless, for privacy-enhanced ML training, our DiVa architecture demonstrates
		its robustness and wide applicability for both non-private and private SGD training.

\begin{figure}[t!] \centering
\includegraphics[width=0.485\textwidth]{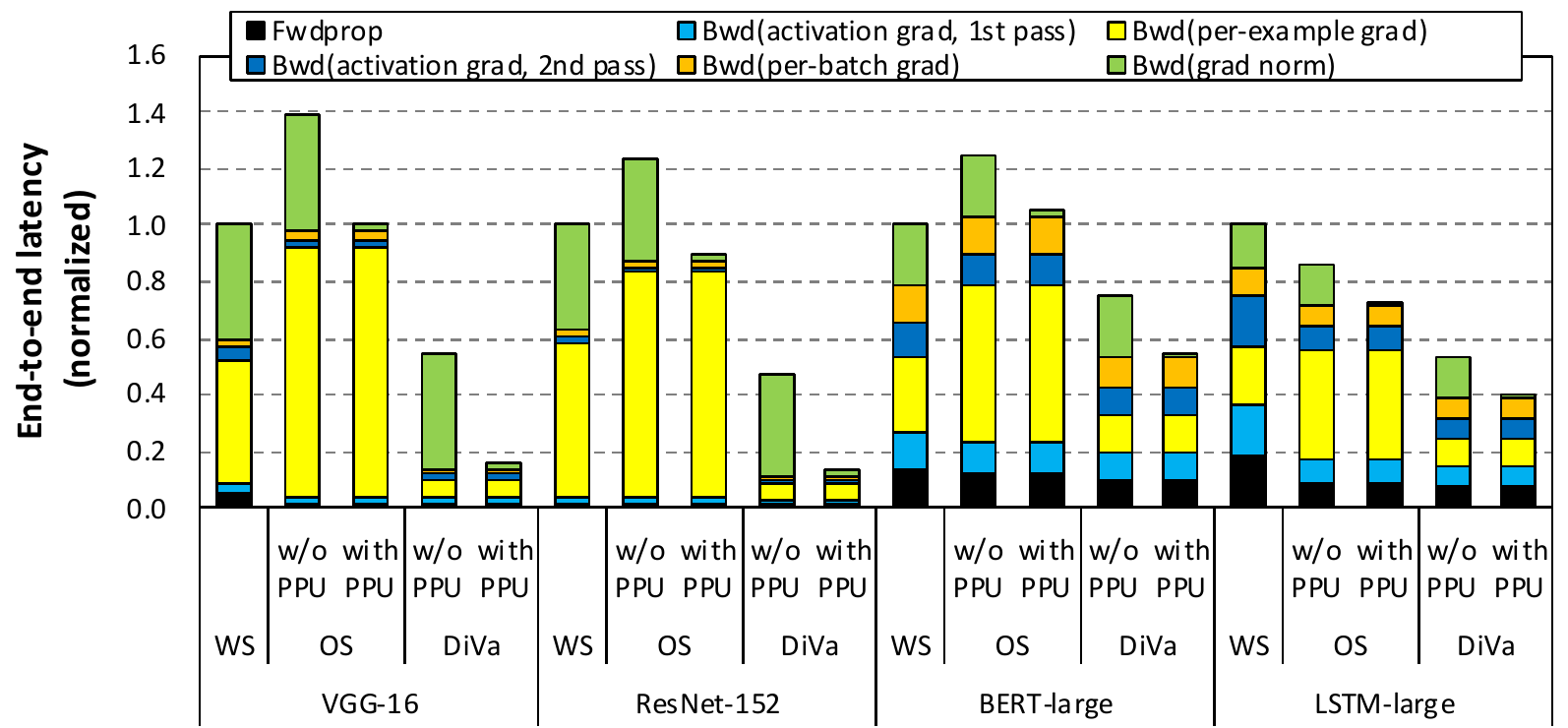}
\caption{
Breakdown of DP training time. Due to space constraints, we only show a subset of our studied DNN models, but the key observations remain intact on models not shown in this figure.
}
\vspace{-0.5em}
\label{fig:eval_latency_breakdown}
\end{figure}

{\bf Latency breakdown.} To better root-cause where \proposed's superior
performance comes from, \fig{fig:eval_latency_breakdown} shows the breakdown of
end-to-end training time. As discussed in
\sect{sect:char_tpu_training_time_breakdown}, derivation of per-example weight
gradients (yellow) and gradient norm (green) causes the biggest performance
degradation. The outer-product based \proposed is the only design point that
successfully addresses the bottlenecks incurred in per-example gradient
derivation, providing an average $7.0\times$ (max $14.6\times$)
	reduction in its latency. Our proposed PPU design also shines with its high
	efficacy, successfully reducing the latency of gradient norm derivation not
	just for \proposed but also for the OS systolic array.

\begin{figure}[t!] \centering
\includegraphics[width=0.485\textwidth]{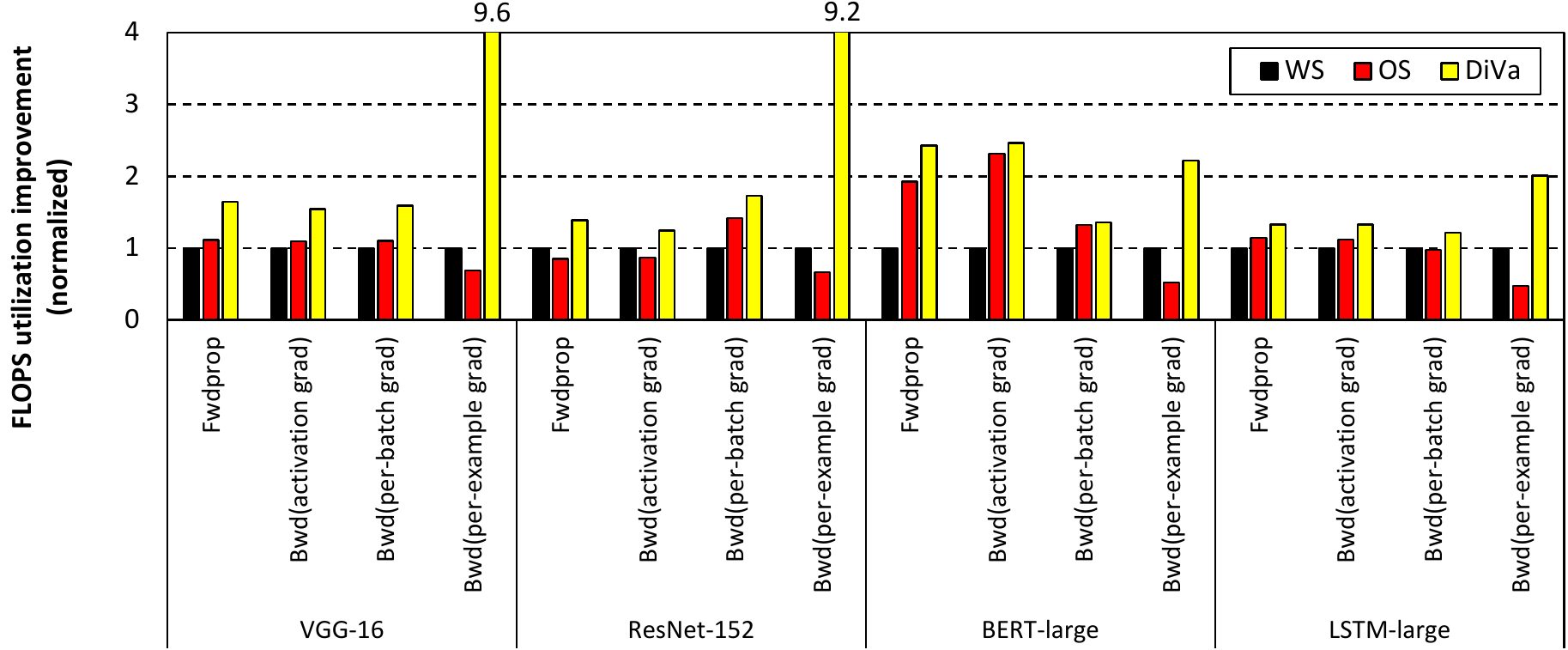}
\caption{
Improvements in FLOPS utilization, i.e., effective computational throughput (normalized to WS systolic array). Similar to \fig{fig:eval_latency_breakdown}, due to space constraints, we discuss a subset of our studied models, but the key observations remain identically over the models not shown in this figure.
}
\vspace{-0.5em}
\label{fig:eval_pe_util}
\end{figure}

{\bf FLOPS utilization.} We highlight \proposed's effectiveness over a different
dimension by presenting the improvements our outer-product dataflow brings about in terms of
FLOPS utilization (\fig{fig:eval_pe_util}). The increase in effective
compute throughput is more
pronounced with CNNs (compared to Transformers/RNNs) as they suffered from more
severe FLOPS underutilization under baseline WS systolic array (see
		\fig{fig:char_tpu_compute_utilization}), achieving an average
$5.5\times$ (max $28.9\times$ in SqueezeNet) improvement in
per-example weight gradient derivations. 
Transformers and RNNs already achieved around $20\%$ effective throughput
even with WS systolic array, so \proposed's benefits are relatively
modest, but still providing a sizable $2.2\times$ average improvement.

\begin{table}[tpb]
	\centering
	\caption{Power, area, and effective throughput (TFLOPS) normalized to power and area.
	}
	\vspace*{-0.5em}
	\scriptsize
	\begin{tabular}{|c|c|c|c|}
	\hline
	\multirow{2}{*}{\bf{}} & \multicolumn{3}{c|}{\bf{GEMM engine}}\\
	\cline{2-4}
	 & Systolic WS & Systolic OS & Outer-product\\
	\hline
	\hline
	Technology & \multicolumn{3}{c|}{Commercial $65$ nm}\\
	\hline
	Clock frequency & \multicolumn{3}{c|}{$940$ MHz} \\
	\hline
	MACs & \multicolumn{3}{c|}{$16,384$ ($128\times$$128$ PEs)} \\
	\hline
	Data type & \multicolumn{3}{c|}{BF16 Mult, FP32 Add}\\
	\hline
	Peak TFLOPS & \multicolumn{3}{c|}{$29.5$}\\
	\hline
	Effective TFLOPS & $1.2$ & $0.9$ & $6.6$ \\ 
	\hline
	Power (Watt) & $13.4$ & $13.6$ & $21.2$ \\
	\hline
	Area (mm$^{2}$) & $68$ & $70$ & $82$\\
	\hline
	Effective TFLOPS/Watt & $0.089$ & $0.070$ & {\bf $0.311$} \\
	\hline
	Effective TFLOPS/mm$^{2}$ & $0.017$ & $0.012$ & {\bf $0.081$} \\
	\hline
	\end{tabular}
	\vspace{-1.5em}
	\label{tab:eval_overhead}
\end{table}

\subsection{Area, Power, and Energy Consumption}
\label{sect:eval_overhead}

{\bf Area and power.} \tab{tab:eval_overhead} summarizes the area and power
overhead of \proposed's GEMM engine and PPU. The
outer-product GEMM engine alone adds $19.6\%$ of area overhead vs. WS systolic
array, with an additional $4.6\%$ overhead with our PPU.  As we target an
accelerator chip with Google TPUv3 level compute throughput and on-chip SRAM
capacity, the chip-wide area is estimated to be $650$ mm$^{2}$ (designed using
		$12$ nm technology)~\cite{jouppi2020domain}.  Consequently, the addition of
\proposed's $17$ (=$85$-$68$) mm$^{2}$ (estimated using $65$ nm standard cell
		library) area costs a chip-wide $0.3\%$
additional area overhead. In terms of power consumption, \proposed's all-to-all
multiplication datapath and PPU adds $10.4$ Watt ($7.8$ (outer-product) + $2.6$ (PPU)) of additional power consumption,
causing a chip-wide $2.3\%$ (=$10.4$/$450$) overhead (i.e., TPUv3's TDP is $450$ W).
In general, such added area and power overheads are
highly reasonable given \proposed's substantial improvement in FLOPS
utilization, achieving $3.5\times$ and $4.6\times$ higher TFLOPS/Watt and
TFLOPS/area than WS, respectively.

\begin{figure}[t!] \centering
\includegraphics[width=0.485\textwidth]{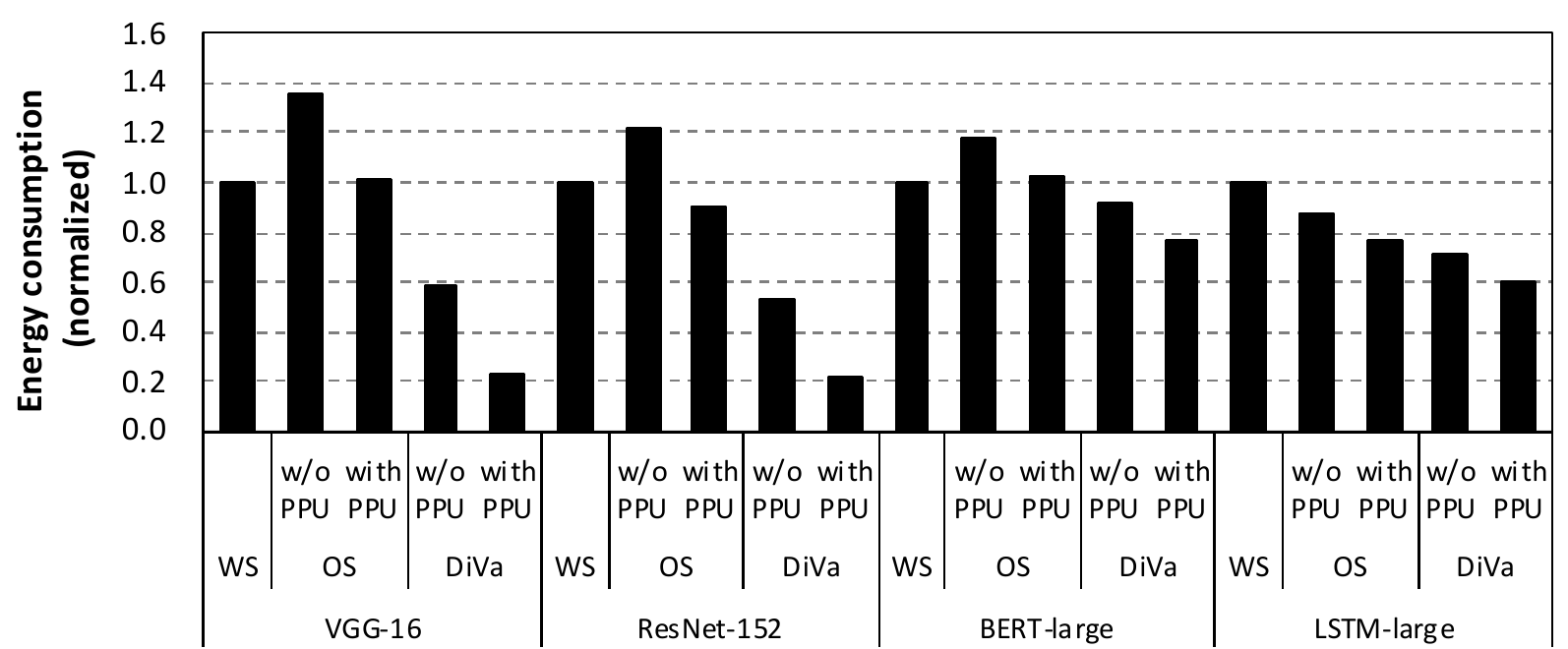}
\caption{
Energy consumption of \proposed (normalized to the WS systolic dataflow).
}
\vspace{-0.5em}
\label{fig:eval_energy}
\end{figure}

{\bf Energy consumption.} Across all the models we study, \proposed provides an
average $2.6\times$ (max $4.6\times$) reduction in energy
		consumption.  Due to space constraints, we show a subset of our studied
		models' chip-wide energy consumption in \fig{fig:eval_energy}. Our analysis
		is based on a $65$ nm technology (i.e.,
		RTL synthesis of \proposed's compute units for power measurement and  
		SRAM energy modeled using CACTI all assume $65$ nm).  As
		depicted, the power overheads of \proposed is outweighed by the
		significant reduction in training time, achieving substantial reduction
		in energy consumption.

\sethlcolor{cyan}
\subsection{Sensitivity}
\label{sect:eval_sensitivity}

This subsection evaluates \proposed's robustness to different model
configurations.  As discussed in \sect{sect:methodology}, DP-SGD for
	computer vision is currently limited to CIFAR-10 level datasets (i.e.,
			$32\times32$ input images), which we assume in our baseline setting. We
		evaluate DiVa's robustness to future larger datasets by increasing the
		image size by $4\times$/$16\times$/$64\times$ (which allows the systolic
				arrays to better populate the PEs for higher throughput), achieving an
		average $3.6\times$/$2.1\times$/$1.7\times$ end-to-end speedup across the
		five CNNs over WS systolic array, respectively.  We also evaluate
		\proposed for Transformers and RNNs with longer input sequence lengths that are
		$2\times$/$4\times$/$8\times$ longer than the baseline $32$ sequence
		length, achieving an average $2.0\times$/$1.6\times$/$1.5\times$ training
		time reduction, respectively.

\sethlcolor{yellow}
\subsection{DiVa vs. GPUs}
\label{sect:eval_diva_vs_gpu}

\begin{figure}[t!] \centering
\includegraphics[width=0.485\textwidth]{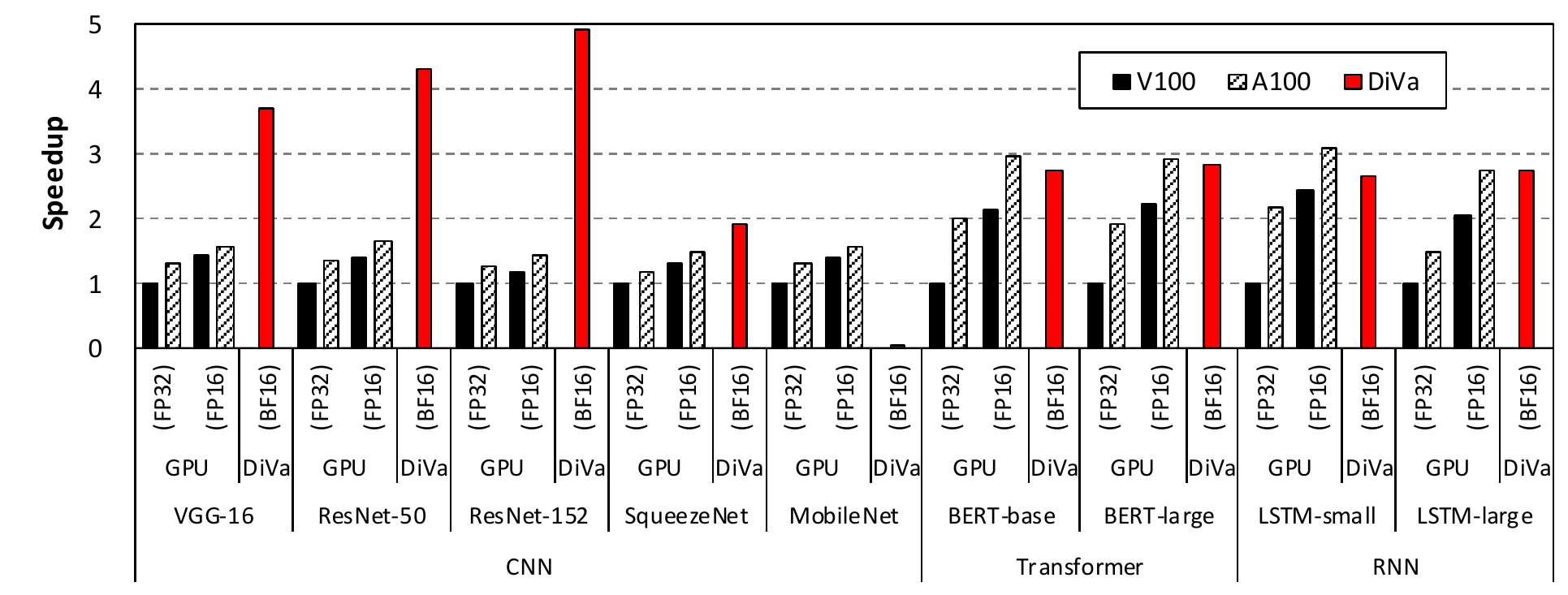}
\caption{
DiVa's speedup vs. NVIDIA's V100 and A100 GPUs. GPU(FP32) and GPU(FP16) refers to V100 and A100 without and with Tensor Cores enabled.
}
\vspace{-0.5em}
\label{fig:eval_gpu_vs_diva}
\end{figure}

While we focused on accelerating DP-SGD over accelerators like Google TPUs, we
also compare \proposed's merits over a GPU system for the completeness of our
study (\fig{fig:eval_gpu_vs_diva}).  We compare DiVa against two GPU
	systems~\cite{volta_v100,ampere_a100} employing NVIDIA's V100 ($32$ GB, $900$ GB/sec of
			bandwidth) and A100 ($40$ GB, $1,555$ GB/sec of bandwidth) running
		JAX enabled with
			auto-vectorization~\cite{jax2018github,SUB_2021,KUR_2022}.  Both V100
			and A100 are evaluated with/without NVIDIA's Tensor Core enabled,
			which provide a sizable difference in its maximum throughput (i.e., $125$
					TFLOPS/$312$ TFLOPS \emph{with} Tensor Cores enabled (FP16) and
					$15.7$ TFLOPS/$19.5$ TFLOPS when disabled (FP32) for
					V100/A100, respectively). For those key GEMM operations that
			constitute DP-SGD's backpropagation bottleneck stages, \proposed
			generally provides superior performance against NVIDIA Tensor Cores
			(FP16), achieving an average $1.2\times$/$1.0\times$ (max
					$4.1\times$/$3.4\times$)  speedup vs. V100/A100,
			respectively, despite having only $23.6\%$/$9.5\%$ of V100 and
				A100's FP16 throughput.  MobileNet is an exception where \proposed
				performs worse than the two GPUs as the GPU seemingly does a better job
				in mapping the small sized GEMMs across its SIMD vector units.
				Nonetheless, recall that \proposed ``only'' comes with $29.5$ TFLOPS,
				unlike V100 and A100's Tensor Cores which contain $4.2\times$
				(=$125$/$29.5$) and $10.6\times$ (=$312$/$29.5$) higher computational
				throughput, respectively, than DiVa.
These results highlight the importance of optimally mapping DP-SGD's MAC operations across
	the computational units (e.g., despite the significantly higher peak TFLOPS of A100 vs. V100, A100
	only achieves incremental speedup comapred to V100), which DiVa's outer-product dataflow 
	demonstrates its efficiency.

\sethlcolor{yellow}					

\sethlcolor{green} 
\section{Related Work}
\label{sect:discussion}

{\bf Outer-product dataflow for GEMM acceleration.} Several recent literature
explored domain-specific architectures for accelerating sparse linear algebra.
Among these, OuterSPACE~\cite{outerspace} and SpArch~\cite{sparch} are two-most
recent studies employing an outer-product dataflow for sparse-sparse GEMMs.
The motivation behind the adoption of outer-product in
OuterSPACE/SpArch is completely different than \proposed as these two studies
seek to reap out opportunities from sparsity.  While the details of the
underlying microarchitecture and its dataflow are not publicly available,
					 Tesla's Full Self-Driving (FSD) computer~\cite{teslamicro,
						 teslahotchips} hints at the adoption of an outer-product dataflow
						 in conducting GEMMs. All of these prior studies assume an
						 \emph{inference} scenario, unlike the training context \proposed
						 is studied over.  More importantly, none of these prior work
						 explores DP training for privacy protection, an important
						 motivation and contribution of our study.

{\bf Accelerators for irregular GEMMs.} Similar to \proposed, SIGMA~\cite{sigma}
seeks to address the PE underutilization issue of systolic arrays in executing
irregular and sparse GEMMs via flexible interconnects and various sparse optimizations.  Unlike the outer-product \proposed, SIGMA employs a SIMD-style inner product
array design, let alone the fact that it is optimized for a non-private SGD
algorithm. Planaria~\cite{planaria} similarly seeks to address the PE underutilization
of systolic arrays for irregular GEMMs via spatially co-locating multiple DNN
models, presenting a dynamically reconfigurable interconnect for better utility
of computation units. 
Again, the focus of these prior studies is different than
our study, rendering the key contribution of our work stands on its own.

{\bf Privacy-preserving ML accelerators.} While not necessarily
exploring differential privacy, there is a body of prior art that seeks to preserve
privacy by adding \emph{security} enhancements. GuardNN~\cite{guardnn} is a DNN accelerator that
employs encryption/decryption and integrity verification for off-chip data movements, enhancing
its security. DarKnight~\cite{darknight} uses a custom data encoding strategy based on matrix masking to enable
input obfuscation. MAXelerator~\cite{maxelerator} proposes a privacy-preserving MAC unit at the circuit-level.
In general, the contributions of \proposed is orthogonal to these prior work.

{\bf DNN dataflows for spatial architectures.}
In this work, we primarily focused on the systolic OS and WS dataflow as assumed
	in our baseline training accelerator. There are however alternative DNN dataflows discussed
	in prior literature, with a particular emphasis on inference deployment scenarios. 
	Eyeriss~\cite{eyeriss_isca}, for instance, argued for
	a row-stationary (RS) dataflow for convolutional neural network inference,
		demonstrating RS's superiority over OS and WS. MAESTRO~\cite{maestro}
explores a data-centric DNN dataflow for inference, presenting an analytical cost model
	to evaluate a target dataflow's latency, throughput, and energy-efficiency.
As discussed in this work, ML training involves the 
derivation of both activation and
	weight gradients, a computation process non-existent in inference.
Therefore, it is unclear how the inference-optimized 
	DNN dataflows
	explored in prior literature~\cite{eyeriss_isca,maestro,gamma,confuciux} can be applied for the backpropagation's gradient derivation.
		Enabling these inference-optimized dataflows to be applicable 
			and optimized for
		training-purposes, let alone DP training, is beyond our scope. Instead, we focused on two most widely
		deployed training-purposed architectures like systolic arrays (i.e., OS and
				WS) and GPUs.

{\bf Multi-tenant ML accelerators for enhanced utilization.}  Aside from novel
DNN dataflows for irregular and/or sparse GEMMs, recent work explored the
possibility of multi-tenant DNN execution as means to improve compute
utilization and throughput for inference. PREMA~\cite{prema} is one of the
first work in this line of research, which employs preemptive multi-tasking to
\emph{temporally} share the ML accelerator. AI-MT~\cite{ai_mt} and
Planaria~\cite{planaria}, on the other hand,  explored \emph{spatial}
multi-tasking to concurrently execute multiple DNN models and better saturate
compute and memory throughput.  Given such, co-locating multiple skinny GEMMs
within the ML accelerator for spatial multi-tasking is an interesting approach
that can potentially lead to higher PE utility in DP-SGD.  However, it is
unclear how such co-location enabled GEMM engine can efficiently handle the
backpropagation stages of deriving both activation and weight gradients (i.e.,
		prior multi-tenant ML accelerators strictly focus on inference, not
		training), a feature naturally supported under the systolic dataflow as
well as our proposed DiVa design.  Optimizing DiVa's spatial array
to handle these cases is beyond the scope of our work and we leave it as future
work.  

\section{Conclusion}
\label{sect:conclusion}

This paper proposed \proposed, an accelerator for differentially private
machine learning training. We first conduct a  workload characterization on a
state-of-the-art DP-SGD algorithm executing over Google TPUs, uncovering its high memory consumption and
low compute utilization issue.  We then utilize the lessons learned from our characterization to
develop a ML accelerator optimized for DP-SGD, employing an
outer-product dataflow augmented with an adder-tree based post-processing unit.
Compared to prior systolic arrays, \proposed provides significant
improvement in compute utilization which allows $3.6\times$ increase in
training throughput.

\section*{Acknowledgment}

This research is partly supported by the National Research Foundation of Korea
(NRF) grant funded by the Korea government(MSIT) (NRF-2021R1A2C2091753), the
Engineering Research Center Program through the NRF funded by the Korean
government MSIT under grant NRF-2018R1A5A1059921, and by Samsung Electronics
Co., Ltd.  We also appreciate the support from the IC Design Education Center
(IDEC), Korea, for the EDA tools.
Minsoo Rhu is the corresponding author.

\bibliographystyle{IEEEtranS}
\bibliography{refs}

\begin{thebibliography}{10}
\providecommand{\url}[1]{#1}
\csname url@samestyle\endcsname
\providecommand{\newblock}{\relax}
\providecommand{\bibinfo}[2]{#2}
\providecommand{\BIBentrySTDinterwordspacing}{\spaceskip=0pt\relax}
\providecommand{\BIBentryALTinterwordstretchfactor}{4}
\providecommand{\BIBentryALTinterwordspacing}{\spaceskip=\fontdimen2\font plus
\BIBentryALTinterwordstretchfactor\fontdimen3\font minus
  \fontdimen4\font\relax}
\providecommand{\BIBforeignlanguage}[2]{{%
\expandafter\ifx\csname l@#1\endcsname\relax
\typeout{** WARNING: IEEEtranS.bst: No hyphenation pattern has been}%
\typeout{** loaded for the language `#1'. Using the pattern for}%
\typeout{** the default language instead.}%
\else
\language=\csname l@#1\endcsname
\fi
#2}}
\providecommand{\BIBdecl}{\relax}
\BIBdecl

\bibitem{tensorflow2015-whitepaper}
M.~Abadi, A.~Agarwal, P.~Barham, E.~Brevdo, Z.~Chen, C.~Citro, G.~S. Corrado,
  A.~Davis, J.~Dean, M.~Devin, S.~Ghemawat, I.~Goodfellow, A.~Harp, G.~Irving,
  M.~Isard, Y.~Jia, R.~Jozefowicz, L.~Kaiser, M.~Kudlur, J.~Levenberg,
  D.~Man\'{e}, R.~Monga, S.~Moore, D.~Murray, C.~Olah, M.~Schuster, J.~Shlens,
  B.~Steiner, I.~Sutskever, K.~Talwar, P.~Tucker, V.~Vanhoucke, V.~Vasudevan,
  F.~Vi\'{e}gas, O.~Vinyals, P.~Warden, M.~Wattenberg, M.~Wicke, Y.~Yu, and
  X.~Zheng, ``{ {TensorFlow}: Large-Scale Machine Learning on Heterogeneous
  Systems},'' \url{https://www.tensorflow.org/}, 2015.

\bibitem{dpsgd}
M.~Abadi, A.~Chu, I.~Goodfellow, H.~B. McMahan, I.~Mironov, K.~Talwar, and
  L.~Zhang, ``{Deep Learning with Differential Privacy},'' in
  \emph{{Proceedings of the ACM SIGSAC Conference on Computer and
  Communications Security (CCS)}}, 2016.

\bibitem{cnvlutin}
J.~Albericio, P.~Judd, T.~Hetherington, T.~Aamodt, N.~E. Jerger, and
  A.~Moshovos, ``{Cnvlutin: Ineffectual-Neuron-Free Deep Convolutional Neural
  Network Computing},'' in \emph{Proceedings of the International Symposium on
  Computer Architecture (ISCA)}, 2016.

\bibitem{anil2021large}
R.~Anil, B.~Ghazi, V.~Gupta, R.~Kumar, and P.~Manurangsi, ``{Large-Scale
  Differentially Private {BERT}},'' in \emph{{arxiv.org}}, 2021.

\bibitem{differential2017learning}
Apple, ``{Learning with Privacy at Scale},''
  \url{https://docs-assets.developer.apple.com/ml-research/papers/learning-with-privacy-at-scale.pdf},
  2017.

\bibitem{ai_mt}
E.~Baek, D.~Kwon, and J.~Kim, ``{A Multi-Neural Network Acceleration
  Architecture},'' in \emph{Proceedings of the International Symposium on
  Computer Architecture (ISCA)}, 2020.

\bibitem{teslahotchips}
P.~Bannon, G.~Venkataramanan, D.~D. Sarma, and E.~Talpes, ``{Computer and
  Redundancy Solution for the Full Self-Driving Computer},'' in \emph{Hot
  Chips: A Symposium on High Performance Chips}, 2019.

\bibitem{bassily2014private}
R.~Bassily, A.~Smith, and A.~Thakurta, ``{Private Empirical Risk Minimization:
  Efficient Algorithms and Tight Error Bounds},'' in \emph{IEEE Annual
  Symposium on Foundations of Computer Science}, 2014.

\bibitem{blatt2020secure}
M.~Blatt, A.~Gusev, Y.~Polyakov, and S.~Goldwasser, ``{Secure Large-Scale
  Genome-wide Association Studies Using Homomorphic Encryption},'' in
  \emph{Proceedings of the National Academy of Sciences}, 2020.

\bibitem{jax2018github}
J.~Bradbury, R.~Frostig, P.~Hawkins, M.~J. Johnson, C.~Leary, D.~Maclaurin,
  G.~Necula, A.~Paszke, J.~Vander{P}las, S.~Wanderman-{M}ilne, and Q.~Zhang,
  ``{{JAX}: Composable Transformations of {P}ython+{N}um{P}y Programs},''
  http://github.com/google/jax, 2022.

\bibitem{brown2020gpt3}
T.~Brown, B.~Mann, N.~Ryder, M.~Subbiah, J.~D. Kaplan, P.~Dhariwal,
  A.~Neelakantan, P.~Shyam, G.~Sastry, A.~Askell \emph{et~al.}, ``{Language
  Models are Few-Shot Learners},'' in \emph{Proceedings of the International
  Conference on Neural Information Processing Systems (NIPS)}, 2020.

\bibitem{nicholas2020blog}
N.~Carlini, ``{Privacy Considerations in Large Language Models},''
  \url{https://ai.googleblog.com/2020/12/privacy-considerations-in-large.html},
  2020.

\bibitem{nicholas2021extracting}
N.~Carlini, F.~Tram{\`e}r, E.~Wallace, M.~Jagielski, A.~Herbert-Voss, K.~Lee,
  A.~Roberts, T.~Brown, D.~Song, {\'U}.~Erlingsson, A.~Oprea, and C.~Raffel,
  ``{Extracting Training Data from Large Language Models},'' in \emph{{USENIX
  Security Symposium (USENIX Security)}}, 2021.

\bibitem{diannao}
T.~Chen, Z.~Du, N.~Sun, J.~Wang, C.~Wu, Y.~Chen, and O.~Temam, ``{DianNao: A
  Small-Footprint High-Throughput Accelerator for Ubiquitous
  Machine-learning},'' in \emph{Proceedings of the International Conference on
  Architectural Support for Programming Languages and Operation Systems
  (ASPLOS)}, 2014.

\bibitem{eyeriss_isca}
Y.~Chen, J.~Emer, and V.~Sze, ``{Eyeriss: A Spatial Architecture for
  Energy-Efficient Dataflow for Convolutional Neural Networks},'' in
  \emph{Proceedings of the International Symposium on Computer Architecture
  (ISCA)}, June 2016.

\bibitem{eyeriss}
Y.~Chen, T.~Krishna, J.~Emer, and V.~Sze, ``{Eyeriss: An Energy-Efficient
  Reconfigurable Accelerator for Deep Convolutional Neural Networks},'' in
  \emph{Proceedings of the International Solid State Circuits Conference
  (ISSCC)}, 2016.

\bibitem{dadiannao}
Y.~Chen, T.~Luo, S.~Liu, S.~Zhang, L.~He, J.~Wang, L.~Li, T.~Chen, Z.~Xu,
  N.~Sun, and O.~Temam, ``{DaDianNao: A Machine-Learning Supercomputer},'' in
  \emph{Proceedings of the International Symposium on Microarchitecture
  (MICRO)}, 2014.

\bibitem{chetlur2014cudnn}
S.~Chetlur, C.~Woolley, P.~Vandermersch, J.~Cohen, J.~Tran, B.~Catanzaro, and
  E.~Shelhamer, ``{cuDNN: Efficient Primitives for Deep Learning},'' in
  \emph{{arxiv.org}}, 2014.

\bibitem{prema}
Y.~Choi and M.~Rhu, ``{PREMA: A Predictive Multi-task Scheduling Algorithm For
  Preemptible Neural Processing Units},'' in \emph{Proceedings of the
  International Symposium on High-Performance Computer Architecture (HPCA)},
  2020.

\bibitem{bert}
J.~Devlin, M.-W. Chang, K.~Lee, and K.~Toutanova, ``{BERT: Pre-training of Deep
  Bidirectional Transformers for Language Understanding},'' in
  \emph{{arxiv.org}}, 2018.

\bibitem{diethe2020preserving}
T.~Diethe, O.~Feyisetan, B.~Balle, and T.~Drake, ``{Preserving Privacy in
  Analyses of Textual Data},'' in \emph{Proceedings of the International
  Conference on Web Search and Data Mining}, 2020.

\bibitem{dong2017dropping}
H.~Dong, C.~Wu, Z.~Wei, and Y.~Guo, ``{Dropping Activation Outputs With
  Localized First-Layer Deep Network for Enhancing User Privacy and Data
  Security},'' in \emph{IEEE Transactions on Information Forensics and
  Security}, 2018.

\bibitem{dwork2006dp}
C.~Dwork, ``{Differential Privacy},'' in \emph{Automata, Languages and
  Programming}, 2006.

\bibitem{dwork2008differential}
C.~Dwork, ``{Differential Privacy: A Survey of Results},'' in \emph{Theory and
  Applications of Models of Computation}, 2008.

\bibitem{dwork2014algorithmic}
C.~Dwork and A.~Roth, ``{The Algorithmic Foundations of Differential
  Privacy},'' in \emph{Found. Trends Theor. Comput. Sci.}, 2014.

\bibitem{jakob2016learning}
J.~N. Foerster, Y.~M. Assael, N.~de~Freitas, and S.~Whiteson, ``{Learning to
  Communicate with Deep Multi-Agent Reinforcement Learning},'' in
  \emph{Proceedings of the International Conference on Neural Information
  Processing Systems (NIPS)}, 2016.

\bibitem{fredrikson2015model}
M.~Fredrikson, S.~Jha, and T.~Ristenpart, ``{Model Inversion Attacks That
  Exploit Confidence Information and Basic Countermeasures},'' in
  \emph{{Proceedings of the ACM SIGSAC Conference on Computer and
  Communications Security (CCS)}}, 2015.

\bibitem{planaria}
S.~Ghodrati, B.~H. Ahn, J.~K. Kim, S.~Kinzer, B.~R. Yatham, N.~Alla, H.~Sharma,
  M.~Alian, E.~Ebrahimi, N.~S. Kim \emph{et~al.}, ``{Planaria: Dynamic
  Architecture Fission for Spatial Multi-Tenant Acceleration of Deep Neural
  Networks},'' in \emph{Proceedings of the International Symposium on
  Microarchitecture (MICRO)}, 2020.

\bibitem{cloud_tpu}
Google, ``{Cloud TPU},'' \url{https://cloud.google.com/tpu}, 2018.

\bibitem{resnet}
S.~Gross and M.~Wilber, ``{Training and Investigating Residual Nets},'' in
  \emph{{arxiv.org}}, 2016.

\bibitem{song:2015:eie}
S.~Han, X.~Liu, H.~Mao, J.~Pu, A.~Pedram, M.~Horowitz, and W.~Dally, ``{EIE:
  Efficient Inference Engine on Compressed Deep Neural Network},'' in
  \emph{Proceedings of the International Symposium on Computer Architecture
  (ISCA)}, June 2016.

\bibitem{darknight}
H.~Hashemi, Y.~Wang, and M.~Annavaram, ``{DarKnight: An Accelerated Framework
  for Privacy and Integrity Preserving Deep Learning Using Trusted Hardware},''
  in \emph{Proceedings of the International Symposium on Microarchitecture
  (MICRO)}, 2021.

\bibitem{hayes2010logan}
J.~Hayes, L.~Melis, G.~Danezis, and E.~De~Cristofaro, ``{LOGAN: Membership
  Inference Attacks Against Generative Models},'' in \emph{{Proceedings on
  Privacy Enhancing Technologies (PoPET)}}, 2019.

\bibitem{heaton2017deep}
J.~B. Heaton, N.~G. Polson, and J.~H. Witte, ``{Deep Learning for Finance: Deep
  Portfolios},'' in \emph{Applied Stochastic Models in Business and Industry},
  2017.

\bibitem{lstm}
S.~Hochreiter and J.~Schmidhuber, ``{Long Short Term Memory},'' in \emph{Neural
  Computation}, 1997.

\bibitem{hoory2021learning}
S.~Hoory, A.~Feder, A.~Tendler, S.~Erell, A.~Cohen, I.~Laish, H.~Nakhost,
  U.~Stemmer, A.~Benjamini, A.~Hassidim, and Y.~Matias, ``{Learning and
  Evaluating a Differentially Private Pre-trained Language Model},'' in
  \emph{Findings of the Association for Computational Linguistics: EMNLP},
  2021.

\bibitem{horowitz:isscc}
M.~Horowitz, ``{Computing's Energy Problem (and What We Can Do About It)},'' in
  \emph{Proceedings of the International Solid State Circuits Conference
  (ISSCC)}, 2014.

\bibitem{mobilenet}
A.~G. Howard, M.~Zhu, B.~Chen, D.~Kalenichenko, W.~Wang, T.~Weyand,
  M.~Andreetto, and H.~Adam, ``{MobileNets: Efficient Convolutional Neural
  Networks for Mobile Vision Applications},'' \emph{arXiv preprint
  arXiv:1704.04861}, 2017.

\bibitem{guardnn}
W.~Hua, M.~Umar, Z.~Zhang, and G.~E. Suh, ``{GuardNN: Secure DNN Accelerator
  for Privacy-Preserving Deep Learning},'' in \emph{{arxiv.org}}, 2020.

\bibitem{maxelerator}
S.~U. Hussain, B.~D. Rouhani, M.~Ghasemzadeh, and F.~Koushanfar,
  ``{MAXelerator: FPGA Accelerator for Privacy Preserving Multiply-Accumulate
  (MAC) on Cloud Servers},'' in \emph{Design Automation Conference (DAC)},
  2018.

\bibitem{centaur}
R.~Hwang, T.~Kim, Y.~Kwon, and M.~Rhu, ``{Centaur: A Chiplet-Based, Hybrid
  Sparse-Dense Accelerator for Personalized Recommendations},'' in
  \emph{Proceedings of the International Symposium on Computer Architecture
  (ISCA)}, 2020.

\bibitem{neummu}
B.~Hyun, Y.~Kwon, Y.~Choi, J.~Kim, and M.~Rhu, ``{NeuMMU: Architectural Support
  for Efficient Address Translations in Neural Processing Units},'' in
  \emph{Proceedings of the International Conference on Architectural Support
  for Programming Languages and Operation Systems (ASPLOS)}, 2020.

\bibitem{squeezenet}
F.~Iandola, S.~Han, M.~Moskewicz, K.~Ashraf, W.~J. Dally, and K.~Keutzer,
  ``{SqueezeNet: AlexNet-level Accuracy with 50x Fewer Parameters and $<$0.5MB
  Model Size},'' in \emph{{arxiv.org}}, 2016.

\bibitem{jia2014learning}
Y.~Jia, ``{Learning Semantic Image Representations at a Large Scale},'' 2014.

\bibitem{tpu1}
N.~P. Jouppi, C.~Young, N.~Patil, D.~Patterson, G.~Agrawal, R.~Bajwa, S.~Bates,
  S.~Bhatia, N.~Boden, A.~Borchers, R.~Boyle, P.~Cantin, C.~Chao, C.~Clark,
  J.~Coriell, M.~Daley, M.~Dau, J.~Dean, B.~Gelb, T.~V. Ghaemmaghami,
  R.~Gottipati, W.~Gulland, R.~Hagmann, C.~R. Ho, D.~Hogberg, J.~Hu, R.~Hundt,
  D.~Hurt, J.~Ibarz, A.~Jaffey, A.~Jaworski, A.~Kaplan, H.~Khaitan,
  D.~Killebrew, A.~Koch, N.~Kumar, S.~Lacy, J.~Laudon, J.~Law, D.~Le, C.~Leary,
  Z.~Liu, K.~Lucke, A.~Lundin, G.~MacKean, A.~Maggiore, M.~Mahony, K.~Miller,
  R.~Nagarajan, R.~Narayanaswami, R.~Ni, K.~Nix, T.~Norrie, M.~Omernick,
  N.~Penukonda, A.~Phelps, J.~Ross, M.~Ross, A.~Salek, E.~Samadiani, C.~Severn,
  G.~Sizikov, M.~Snelham, J.~Souter, D.~Steinberg, A.~Swing, M.~Tan,
  G.~Thorson, B.~Tian, H.~Toma, E.~Tuttle, V.~Vasudevan, R.~Walter, W.~Wang,
  E.~Wilcox, and D.~H. Yoon, ``{In-Datacenter Performance Analysis of a Tensor
  Processing Unit},'' in \emph{Proceedings of the International Symposium on
  Computer Architecture (ISCA)}, 2017.

\bibitem{jouppi2021ten}
N.~P. Jouppi, D.~H. Yoon, M.~Ashcraft, M.~Gottscho, T.~B. Jablin, G.~Kurian,
  J.~Laudon, S.~Li, P.~Ma, X.~Ma \emph{et~al.}, ``{Ten Lessons From Three
  Generations Shaped Google's TPUv4i : Industrial Product},'' in
  \emph{Proceedings of the International Symposium on Computer Architecture
  (ISCA)}, 2021.

\bibitem{jouppi2020domain}
N.~P. Jouppi, D.~H. Yoon, G.~Kurian, S.~Li, N.~Patil, J.~Laudon, C.~Young, and
  D.~Patterson, ``{A Domain-Specific Supercomputer for Training Deep Neural
  Networks},'' in \emph{Communications of the ACM}, 2020.

\bibitem{kalamkar2019study}
D.~Kalamkar, D.~Mudigere, N.~Mellempudi, D.~Das, K.~Banerjee, S.~Avancha, D.~T.
  Vooturi, N.~Jammalamadaka, J.~Huang, H.~Yuen \emph{et~al.}, ``{A Study of
  BFLOAT16 for Deep Learning Training},'' in \emph{{arxiv.org}}, 2019.

\bibitem{confuciux}
S.-C. Kao, G.~Jeong, and T.~Krishna, ``{ConfuciuX: Autonomous Hardware Resource
  Assignment for DNN Accelerators using Reinforcement Learning},'' in
  \emph{Proceedings of the International Symposium on Microarchitecture
  (MICRO)}, 2020.

\bibitem{gamma}
S.-C. Kao and T.~Krishna, ``{GAMMA: Automating the HW Mapping of DNN Models on
  Accelerators via Genetic Algorithm},'' in \emph{{Proceedings of the 39th
  International Conference on Computer-Aided Design}}, 2020.

\bibitem{kim2020trim}
B.~Kim, J.~Park, E.~Lee, M.~Rhu, and J.~H. Ahn, ``{TRiM: Tensor Reduction in
  Memory},'' in \emph{IEEE Computer Architecture Letters}, 2020.

\bibitem{neurocube}
D.~Kim, J.~Kung, S.~Chai, S.~Yalamanchili, and S.~Mukhopadhyay, ``{Neurocube: A
  Programmable Digital Neuromorphic Architecture with High-Density 3D
  Memory},'' in \emph{Proceedings of the International Symposium on Computer
  Architecture (ISCA)}, 2016.

\bibitem{KUR_2022}
A.~Kurakin, S.~Song, S.~Chien, R.~Geambasu, A.~Terzis, and A.~Thakurta,
  ``{Toward Training at ImageNet Scale with Differential Privacy},'' in
  \emph{{arxiv.org}}, 2022.

\bibitem{maestro}
H.~Kwon, P.~Chatarasi, V.~Sarkar, T.~Krishna, M.~Pellauer, and A.~Parashar,
  ``{MAESTRO: A Data-Centric Approach to Understand Reuse, Performance, and
  Hardware Cost of DNN Mappings},'' in \emph{{IEEE Micro}}, 2020.

\bibitem{maeri}
H.~Kwon, A.~Samajdar, and T.~Krishna, ``{MAERI: Enabling Flexible Dataflow
  Mapping over DNN Accelerators via Programmable Interconnects},'' in
  \emph{Proceedings of the International Conference on Architectural Support
  for Programming Languages and Operation Systems (ASPLOS)}, 2018.

\bibitem{tensordimm}
Y.~Kwon, Y.~Lee, and M.~Rhu, ``{TensorDIMM: A Practical Near-Memory Processing
  Architecture for Embeddings and Tensor Operations in Deep Learning},'' in
  \emph{Proceedings of the International Symposium on Microarchitecture
  (MICRO)}, 2019.

\bibitem{tensorcasting}
Y.~Kwon, Y.~Lee, and M.~Rhu, ``{Tensor Casting: Co-Designing
  Algorithm-Architecture for Personalized Recommendation Training},'' in
  \emph{Proceedings of the International Symposium on High-Performance Computer
  Architecture (HPCA)}, 2021.

\bibitem{mcdla:cal}
Y.~Kwon and M.~Rhu, ``{{A Case for Memory-Centric HPC System Architecture for
  Training Deep Neural Networks}},'' in \emph{IEEE Computer Architecture
  Letters}, 2018.

\bibitem{mcdla}
Y.~Kwon and M.~Rhu, ``{Beyond the Memory Wall: A Case for Memory-Centric HPC
  System for Deep Learning},'' in \emph{Proceedings of the International
  Symposium on Microarchitecture (MICRO)}, 2018.

\bibitem{kwon:2019:disagg}
Y.~Kwon and M.~Rhu, ``{A Disaggregated Memory System for Deep Learning},'' in
  \emph{IEEE Micro}, 2019.

\bibitem{cacti}
H.~Labs, ``{CACTI: An Integrated Cache and Memory Access Time, Cycle Time,
  Area, Leakage, and Dynamic Power Model},''
  \url{http://www.hpl.hp.com/research/cacti/}, 2016.

\bibitem{lecun_gd}
Y.~LeCun, L.~Bottou, Y.~Bengio, and P.~Haffner, ``{Gradient-Based Learning
  Applied to Document Recognition},'' in \emph{Proceedings of the IEEE}, 1998.

\bibitem{lee2021scaling}
J.~Lee and D.~Kifer, ``{Scaling up Differentially Private Deep Learning with
  Fast Per-Example Gradient Clipping},'' in \emph{{Proceedings on Privacy
  Enhancing Technologies (PoPET)}}, 2021.

\bibitem{li2022large}
X.~Li, F.~Tramer, P.~Liang, and T.~Hashimoto, ``{Large Language Models Can Be
  Strong Differentially Private Learners},'' in \emph{{Proceedings of the
  International Conference on Learning Representations (ICLR)}}, 2022.

\bibitem{cambricon}
S.~Liu, Z.~Du, J.~Tao, D.~Han, T.~Luo, Y.~Xie, Y.~Chen, and T.~Chen,
  ``{Cambricon: An Instruction Set Architecture for Neural Networks},'' in
  \emph{Proceedings of the International Symposium on Computer Architecture
  (ISCA)}, 2016.

\bibitem{nasr2019comprehensive}
M.~Nasr, R.~Shokri, and A.~Houmansadr, ``{Comprehensive Privacy Analysis of
  Deep Learning: Passive and Active White-box Inference Attacks Against
  Centralized and Federated Learning},'' in \emph{{IEEE Symposium on Security
  and Privacy (SP)}}, 2019.

\bibitem{norrie2020google}
T.~Norrie, N.~Patil, D.~H. Yoon, G.~Kurian, S.~Li, J.~Laudon, C.~Young, N.~P.
  Jouppi, and D.~A. Patterson, ``{Google's Training Chips Revealed: TPUv2 and
  TPUv3.}'' in \emph{Hot Chips: A Symposium on High Performance Chips}, 2020.

\bibitem{volta_v100}
NVIDIA, ``{NVIDIA Tesla V100},'' 2018.

\bibitem{ampere_a100}
NVIDIA, ``{NVIDIA A100},'' 2020.

\bibitem{charlstmclassification}
Opacus,
  \url{https://github.com/pytorch/opacus/blob/main/examples/char-lstm-classification.py},
  2020.

\bibitem{outerspace}
S.~Pal, J.~Beaumont, D.-H. Park, A.~Amarnath, S.~Feng, C.~Chakrabarti, H.-S.
  Kim, D.~Blaauw, T.~Mudge, and R.~Dreslinski, ``{OuterSPACE: An Outer Product
  Based Sparse Matrix Multiplication Accelerator},'' in \emph{Proceedings of
  the International Symposium on High-Performance Computer Architecture
  (HPCA)}, 2018.

\bibitem{pate}
N.~Papernot, M.~Abadi, U.~Erlingsson, I.~Goodfellow, and K.~Talwar,
  ``{Semi-supervised Knowledge Transfer for Deep Learning from Private Training
  Data},'' in \emph{{Proceedings of the International Conference on Learning
  Representations (ICLR)}}, 2017.

\bibitem{tfprivacy}
N.~Papernot, A.~Galen, and S.~Chien, ``{Tensorflow Privacy},''
  \url{https://github.com/tensorflow/privacy}, 2022.

\bibitem{scnn}
A.~Parashar, M.~Rhu, A.~Mukkara, A.~Puglielli, R.~Venkatesan, B.~Khailany,
  J.~Emer, S.~W. Keckler, and W.~J. Dally, ``{SCNN: An Accelerator for
  Compressed-sparse Convolutional Neural Networks},'' in \emph{Proceedings of
  the International Symposium on Computer Architecture (ISCA)}, 2017.

\bibitem{trim}
J.~Park, B.~Kim, S.~Yun, E.~Lee, M.~Rhu, and J.~H. Ahn, ``{TRiM: Enhancing
  Processor-Memory Interfaces with Scalable Tensor Reduction in Memory},'' in
  \emph{Proceedings of the International Symposium on Microarchitecture
  (MICRO)}, 2021.

\bibitem{sigma}
E.~Qin, A.~Samajdar, H.~Kwon, V.~Nadella, S.~Srinivasan, D.~Das, B.~Kaul, and
  T.~Krishna, ``{SIGMA: A Sparse and Irregular GEMM Accelerator with Flexible
  Interconnects for DNN Training},'' in \emph{Proceedings of the International
  Symposium on High-Performance Computer Architecture (HPCA)}, 2020.

\bibitem{gpt2}
A.~Radford, J.~Wu, R.~Child, D.~Luan, D.~Amodei, and I.~Sutskever, ``{Language
  Models Are Unsupervised Multitask Learners},'' in \emph{OpenAI Blog}, 2019.

\bibitem{rhu:2016:vdnn}
M.~Rhu, N.~Gimelshein, J.~Clemons, A.~Zulfiqar, and S.~W. Keckler, ``{vDNN:
  Virtualized Deep Neural Networks for Scalable, Memory-Efficient Neural
  Network Design},'' in \emph{Proceedings of the International Symposium on
  Microarchitecture (MICRO)}, October 2016.

\bibitem{rhu:2018:cdma}
M.~Rhu, M.~O'Connor, N.~Chatterjee, J.~Pool, Y.~Kwon, and S.~W. Keckler,
  ``{Compressing DMA Engine: Leveraging Activation Sparsity for Training Deep
  Neural Networks},'' in \emph{Proceedings of the International Symposium on
  High-Performance Computer Architecture (HPCA)}, February 2018.

\bibitem{tpu_patent3}
J.~Ross, ``{Prefetching Weights for Use in a Neural Network Processor},''
  Patent, 05 2015, uS 9805304B2.

\bibitem{tpu_patent1}
J.~Ross, N.~Jouppi, A.~Phelps, R.~Young, T.~Norrie, G.~Thorson, and D.~Luu,
  ``{Neural Network Processor},'' Patent, 05 2015, uS 9747546B2.

\bibitem{tpu_patent2}
J.~Ross and A.~Phelps, ``{Computing Convolutions Using a Neural Network
  Processor},'' Patent, 05 2015, uS 9697463B2.

\bibitem{tpu_patent4}
J.~Ross and G.~Thorson, ``{Rotating Data for Neural Network Computations},''
  Patent, 05 2015, uS 9747548B2.

\bibitem{shazeer2017moe}
N.~Shazeer, A.~Mirhoseini, K.~Maziarz, A.~Davis, Q.~Le, G.~Hinton, and J.~Dean,
  ``{Outrageously Large Neural Networks: The Sparsely-Gated Mixture-of-Experts
  Layer},'' in \emph{{Proceedings of the International Conference on Learning
  Representations (ICLR)}}, 2017.

\bibitem{shokri2015ppdl}
R.~Shokri and V.~Shmatikov, ``{Privacy-Preserving Deep Learning},'' in
  \emph{{Proceedings of the ACM SIGSAC Conference on Computer and
  Communications Security (CCS)}}, 2015.

\bibitem{shokri2017membership}
R.~Shokri, M.~Stronati, C.~Song, and V.~Shmatikov, ``{Membership Inference
  Attacks Against Machine Learning Models},'' in \emph{{IEEE Symposium on
  Security and Privacy (SP)}}, 2017.

\bibitem{alphago}
D.~Silver, A.~Huang, C.~J. Maddison, A.~Guez, L.~Sifre, G.~van~den Driessche,
  J.~Schrittwieser, I.~Antonoglou, V.~Panneershelvam, M.~Lanctot, S.~Dieleman,
  D.~Grewe, J.~Nham, N.~Kalchbrenner, I.~Sutskever, T.~Lillicrap, M.~Leach,
  K.~Kavukcuoglu, T.~Graepel, and D.~Hassabis, ``{Mastering the Game of {Go}
  with Deep Neural Networks and Tree Search},'' in \emph{Nature}, 2016.

\bibitem{vggnet}
K.~Simonyan and A.~Zisserman, ``{Very Deep Convolutional Networks for
  Large-Scale Image Recognition},'' in \emph{{arxiv.org}}, 2014.

\bibitem{smith2022using}
S.~Smith, M.~Patwary, B.~Norick, P.~LeGresley, S.~Rajbhandari, J.~Casper,
  Z.~Liu, S.~Prabhumoye, G.~Zerveas, V.~Korthikanti \emph{et~al.}, ``{Using
  DeepSpeed and Megatron to Train Megatron-Turing NLG 530B, A Large-Scale
  Generative Language Model},'' in \emph{{arxiv.org}}, 2022.

\bibitem{SUB_2021}
P.~Subramani, N.~Vadivelu, and G.~Kamath, ``{Enabling Fast Differentially
  Private SGD via Just-in-Time Compilation and Vectorization},'' in
  \emph{Proceedings of the International Conference on Neural Information
  Processing Systems (NIPS)}, 2021.

\bibitem{teslamicro}
E.~Talpes, D.~D. Sarma, G.~Venkataramanan, P.~Bannon, B.~McGee, B.~Floering,
  A.~Jalote, C.~Hsiong, S.~Arora, A.~Gorti, and G.~S. Sachdev, ``{Compute
  Solution for Tesla's Full Self-Driving Computer},'' in \emph{{IEEE Micro}},
  2020.

\bibitem{yu2022differentially}
D.~Yu, S.~Naik, A.~Backurs, S.~Gopi, H.~A. Inan, G.~Kamath, J.~Kulkarni, Y.~T.
  Lee, A.~Manoel, L.~Wutschitz, S.~Yekhanin, and H.~Zhang, ``{Differentially
  Private Fine-tuning of Language Models},'' in \emph{{Proceedings of the
  International Conference on Learning Representations (ICLR)}}, 2022.

\bibitem{sparch}
Z.~Zhang, H.~Wang, S.~Han, and W.~J. Dally, ``{SpArch: Efficient Architecture
  for Sparse Matrix Multiplication},'' in \emph{Proceedings of the
  International Symposium on High-Performance Computer Architecture (HPCA)},
  2020.

\end{thebibliography}

\end{document}